\documentclass[aps,prd,longbibliography,nofootinbib,notitlepage,superscriptaddress]{revtex4-1}
\usepackage{graphicx}
\usepackage{amsfonts,amsthm}
\usepackage{amsmath,amssymb}
\usepackage{bm}
\usepackage[utf8]{inputenc}
\usepackage[usenames,dvipsnames]{xcolor}

\newcommand{\be}{\begin{equation}}
\newcommand{\ee}{\end{equation}}
\newcommand{\bea}{\begin{eqnarray}}
\newcommand{\eea}{\end{eqnarray}}
\newcommand{\dst}{\displaystyle}
\newcommand{\lr}[1]{ \langle #1 \rangle}

\newcommand{\fr}[2]{\frac{{\dst #1}}{{\dst #2}}}

\newcommand{\bk}{{\bf k}}
\newcommand{\bK}{{\bf K}}

\newcommand{\br}{{\bf r}}
\newcommand{\bv}{{\bf v}}

\newcommand{\GeV}{\mathrm{GeV}}

\def\lsim{\mathrel{\rlap{\lower4pt\hbox{\hskip1pt$\sim$}}
    \raise1pt\hbox{$<$}}}         
\def\gsim{\mathrel{\rlap{\lower4pt\hbox{\hskip1pt$\sim$}}
    \raise1pt\hbox{$>$}}}         

    \usepackage{bbold}

\graphicspath{{./figs/}}

\newcommand{\de}{\delta}

\newcommand{\si}{\sigma}

\begin{document}
	\title{
		{\normalsize \hfill CFTP/19-031} \\*[7mm]
		Kinematic surprises in twisted particle collisions}
	\author{Igor P. Ivanov}
	\email{igor.ivanov@tecnico.ulisboa.pt}
	\affiliation{CFTP, Instituto Superior Tecnico, Universidade de Lisboa, Lisbon 1049-001, Portugal}
	\author{Nikolai Korchagin}
	\email{korchagin@impcas.ac.cn}
	\affiliation{Institute of Modern Physics, Chinese Academy of Sciences, Lanzhou 730000, China}
	\author{Alexandr Pimikov}
	\email{pimikov@mail.ru}
	\affiliation{Institute of Modern Physics, Chinese Academy of Sciences, Lanzhou 730000, China}
	\affiliation{Research Institute of Physics, Southern Federal University, Rostov-na-Donu 344090, Russia}
	\author{Pengming Zhang}
	\email{zhangpm5@mail.sysu.edu.cn}
	\affiliation{School of Physics and Astronomy, Sun Yat-sen University, Zhuhai 519082, China}



\begin{abstract}
``Twisted particles'' refer to non-plane-wave states of photons, electrons, hadrons, or any other particle 
which carry non-zero, adjustable orbital angular momentum with respect to their average propagation direction.
Twisted photons and electrons have already been experimentally demonstrated, 
and one can expect creation of twisted states of other particles in future.
Such states can be brought in collisions, offering a completely new degree of freedom
in collider experiments and, especially, a novel tool for hadronic physics. 
We recently showed that $2 \to 1$ processes with two twisted particles
such as resonance production in twisted $e^+e^-$ annihilation give access to observables 
which are difficult or impossible to probe in the usual plane-wave collisions.
In this paper, we discuss in detail surprising kinematic features of this process,
focusing on spinless particle annihilation.
They include (1) a new dimension in the final momentum space available in twisted annihilation,
(2) interference fringes emerging in the cross section as a function of the total energy,
and (3) the built-in mass spectrometric capability of this process,
that is, simultaneous production and automatic angular separation of several resonances
with different masses in monochromatic twisted particle annihilation experiment running at fixed energy.
All these features cannot be obtained in the usual plane wave collision setting.
\end{abstract}


\maketitle

\section{Introduction}

High-energy collisions of particles offer direct access to fundamental laws of subatomic world.
In these collisions, new particles may be produced, and their production cross sections
as well as kinematic distributions reveal both the properties of the fundamental interactions and 
the structure of composite particles, such as hadrons.
However, deciphering this information from the outcomes of individual collision events is not an easy task,
partly because of the event-by-event fluctuations of these outcomes.
Moreover, even when the final state is measured by the detector, there can exist 
several pathways (such as production and decay on intermediate hadronic resonances) 
which link the same initial and final states and whose contributions interfere in the scattering amplitude.
In order to disentangle them, one plots the cross section 
as a function of the final state momenta or the initial collision parameters
such as the total center of mass energy and, in some cases, the polarization of initial particles.
Any new experimental tool which would facilitate this task is always welcome.
 
In the recent paper \cite{recent} we proposed a novel way to gain insight into collision processes
and, in particular, into their spin and parity dependences by colliding initial particles prepared 
in the so-called twisted state.
A twisted particle, be it a photon, an electron, a hadron, or anything else,
is a wave packet with helicoidal wave fronts.
Such a wave packet propagates, as a whole, in a certain direction and also carries 
an adjustable orbital angular momentum (OAM) projection with respect to that direction.
Twisted photons \cite{Molina-Terriza:2007,Paggett:2017,Knyazev:2019} and electrons \cite{Bliokh:2007ec,Bliokh:2017uvr,Lloyd:2017} 
have already been demonstrated experimentally. Although their energies
are still too low to be of interest to particle physics community, there exist suggestions
of how to upscatter them into the GeV energy range \cite{Jentschura:2010ap,Jentschura:2011ih}
and how to steer twisted electrons in accelerators without destroying their OAM \cite{Silenko:2019ziz}.
We argued in \cite{recent} that, by making proper use of this new degree of freedom for both initial state particles,
one can access in unpolarized inclusive processes certain observables, which are either unmeasurable
in the plane wave collisions or require control over the polarization state of the 
initial particles or an elaborate analysis of the final state distributions.
In short, twisted particle collisions offer novel ways to look into the scattering process.

This suggestion is not completely new.
The first analysis of kinematic features which arise in collision of two twisted particles\footnote{There is an even larger literature 
on scattering processes in which only one of the initial particles is twisted, see for instance reviews \cite{Knyazev:2019,Bliokh:2017uvr}
and the recent calculation of twisted neutron scattering on nuclei \cite{Afanasev:2019rlo}.
Although this collision setting is perhaps easier to achieve experimentally,
it does not share the essential novelties offered by two twisted particle collision.
In particular, collisions with only one twisted particle cannot produce interference fringes
discussed below and the observables which rely on them.}
was presented in \cite{Ivanov:2011kk}, soon after three experimental groups, following the suggestions of \cite{Bliokh:2007ec},
experimentally produced moderately relativistic electrons \cite{Uchida:2010,Verbeeck:2010,McMorran:2011}.
In subsequent papers \cite{Ivanov:2012na,Ivanov:2016oue} it was shown that (in)elastic scattering of
two twisted particles provides access to the phase of the overall scattering amplitude,
and in particular, to the Coulomb phase of elastic electron scattering.
The idea that collision of particles in an engineered spatial state can reveal more information on the process 
than the usual plane-wave scattering was explored further in \cite{Karlovets:2016dva,Karlovets:2016jrd}
with various examples of non-plane-wave states.

However, all these works studied the $2\to 2$ scattering processes. What is suggested in \cite{recent}
and will be explored in the present work is to look into an even simpler process $2 \to 1$ 
such as hadronic resonance production in twisted $e^+e^-$ annihilation.
It leads to a surprisingly rich list of new phenomena, many of which do not have plane wave counterparts
and may even seem counter-intuitive.

In the present paper, we will focus on kinematic peculiarities which accompany annihilation of two twisted particles.
They include:
\begin{itemize}
	\item appearance of a new dimension in the final state momentum distribution in the monochromatic twisted particle collisions, 
	\item appearance of interference fringes as demonstrated by the $2\to 1$ production cross section of a single final particle with mass $M$ 
	as a function of the collision energy;
	\item ''built-in mass spectrometry'', that is, simultaneous production of two or more resonances with different masses
	in collisions of monochromatic twisted particles with fixed energies, followed by an automatic angular separation of the produced resonances;
	\item the possibility to selectively enhance or suppress production	of these resonances by adjusting the initial kinematics.
\end{itemize}
All these features offer a surprising degree of control over the produced system, which is yet to be explored.

In this paper, we assume that the two initial twisted particles are spinless, that is, scalar fields;
the analysis of spin and parity physics novelties is postponed to the follow-up paper \cite{spin}.
To simplify the exposition, we will also assume that the two colliding particles are massless; 
inclusion of a finite mass is straightforward.
The kinematic features listed above are universal and will apply to twisted $\gamma\gamma$ collisions,
twisted $e^+e^-$ annihilation, twisted (in)elastic $ep$ scattering, and so on.

In the next section, we will remind the reader of how twisted particles are described and 
present an overview of the generic features of twisted particle collisions.
Then in section~\ref{sec:2to1.scalar.case} we will derive the cross section 
of two twisted particle annihilation, first under the assumption of pure Bessel beams and later
by considering realistic twisted states.
In section~\ref{section-mass-spectrometer} we will describe the built-in mass-spectrometric capability
of twisted particle annihilation.
In all case, we will provide a clear qualitative picture and confirm it with numerical examples.
We will end with conclusions; auxiliary expressions can be found in the appendix. 

Throughout the paper, we will use natural units $\hbar = c = 1$.
Three-dimensional vectors will be denoted by bold symbols,
while the transverse momenta will be labeled by the subscript~$\perp$.

\section{Processes with twisted particles}\label{section-description}

\subsection{Description of twisted scalar particles}\label{subsection-description-scalar}

A Bessel twisted state $|E,\varkappa,m\rangle$ is a solution of the free wave equation 
with a definite energy $E$, longitudinal momentum $k_z$, 
modulus of the transverse momentum $|\bk_\perp|=\varkappa$ 
and a definite $z$-projection of the orbital angular momentum $m$, which must be integer.
When written in cylindric coordinates $\rho, \varphi_r, z$, this solution
has the form
 \be
|E,\varkappa, m\rangle = e^{-i E t + i k_z z} \cdot
\psi_{\varkappa m}(\br_\perp)\,, \quad \psi_{\varkappa m}(\br_\perp) = {e^{i m \varphi_r} \over\sqrt{2\pi}}\sqrt{\varkappa}J_{m}(\varkappa \rho)\,,
 \label{twisted-coordinate}
  \ee
where $J_m(x)$ is the Bessel function. 
This function is normalized according to
\begin{equation}
\int d^2\br_\perp \psi^*_{\varkappa' m'}(\br_\perp) \psi_{\varkappa m}(\br_\perp) 
= \delta(\varkappa-\varkappa')\delta_{m, m'}\,.\label{normalization}
\end{equation}
The azimuthal angle dependence $\propto e^{im\varphi_r}$ is the hallmark of the phase vortex.
A twisted state can be represented as a superposition of plane waves:
 \be
|E,\varkappa,m\rangle = e^{-i E t + i k_z z} \int {d^2 \bk_\perp \over(2\pi)^2}a_{\varkappa m}(\bk_\perp) e^{i\bk_\perp \br_\perp}\,,
 \label{twisted-def}
  \ee
where
 \be
a_{\varkappa m}(\bk_\perp)= (-i)^m e^{im\varphi_k}\sqrt{2\pi \over \varkappa}\; \delta(|\bk_\perp|-\varkappa)\label{a}
  \ee
is the corresponding Fourier amplitude. This expansion can be inverted \cite{Ivanov:2011kk},
which means that twisted states form a complete basis for (transverse) wave functions.

Sometimes a different normalization of $a_{\varkappa m}(\bk_\perp)$ appears in the literature 
on twisted particles, namely, with the coefficient $2\pi/\varkappa$ instead of $\sqrt{2\pi/\varkappa}$.
This is the consequence of a different normalization condition for the coordinate wave function: with or without the prefactor
$2\pi/\varkappa$ in Eq.~\eqref{normalization}. This difference does not change the observables;
one just needs to keep track of the exact normalization choice when calculating the event rate and the flux.

If the above Bessel state describes a particle with mass $\mu$, its energy and momentum are related as
$E^2 = \mu^2 + \varkappa^2 + k_z^2$.
However, the {\em average} momentum of this state $\langle\bk\rangle = (0, 0, k_z)$
does not satisfy this dispersion relation:
\be
E^2 = \mu^2 + \varkappa^2 + \langle\bk\rangle^2 \not = \mu^2 + \langle\bk\rangle^2\,.
\ee
Whether to interpret the quantity $\mu^2 + \varkappa^2$ as a new ``effective mass'' squared,
restoring the usual dispersion relation,
is a matter of terminological convenience and does not change physics.

Just like a plane wave, a pure Bessel state $|E,\varkappa,m\rangle$ with fixed $\varkappa$ 
is non-normalizable in the transverse plane. Therefore, when calculating flux and scattering cross section,
one should regularize the divergent expressions with a finite normalization volume and remove it in the end,
see details in \cite{Jentschura:2010ap,Jentschura:2011ih,Ivanov:2011kk,Karlovets:2012eu}.
This must be done with sufficient care, especially if dealing with several twisted particles
defined with respect to different axes \cite{Ivanov:2011bv}.
A much more physically appealing approach is to use realistic monochromatic beams of finite transverse extent.
Such a beam can be written as a superposition of pure Bessel states 
with equal energies and equal OAMs but with a distribution over $\varkappa$:
\be
|E,\bar\varkappa,\sigma,m\rangle = \int d\varkappa \, f(\varkappa) |E,\varkappa,m\rangle\,,\label{WP}
\ee
with a properly normalized weight function $f(\varkappa)$ peaked at $\bar\varkappa$ and having width $\sigma$. 

In a pure Bessel state, the longitudinal and transverse dynamics factor out.
As a result, one can perform a longitudinal boost, shifting the longitudinal momentum but keeping
the transverse distribution unchanged, and the state will remain monochromatic.
In particular, one can find a reference frame where $k_z = 0$,
representing a cylindric standing wave, non-propagating on average.
For the localized state \eqref{WP}, a longitudinal boost will destroy monochromaticity.
It highlights the fact that, in order to properly defined a realistic twisted state,
one must fix a reference frame and an axis in it.
Eventually, the exact description of such twisted states will depend on future experimental capabilities.

\subsection{Collisions of twisted particles: the broad picture}\label{subsection:broad-picture}

Collisions of twisted particles bring in several new aspects which are absent in the plane wave case.
We find it instructive to first provide a qualitative description of the process before going into detailed calculations.

First, one may consider different collision regimes, with either one or both initial particles in the twisted state.
Also, one has a similar choice when describing final particles either as plane waves or as twisted states.
Different collision regimes were studied already in the first papers \cite{Jentschura:2010ap,Jentschura:2011ih,Ivanov:2011kk};
an overview of results in various schemes was given in \cite{Bliokh:2017uvr}.
Here, we will consider the collision setting in which both initial particles are twisted 
while the final state is described with plane waves.
It is in this regime that one observes the novel phenomenon, interference of two plane-wave scattering amplitudes,
which represents the analogue of Young two-slit experiment in momentum space \cite{Ivanov:2016jzt,Ivanov:2016oue}.
At the same time, one avoids the extremely challenging task of determining whether the produced particle
is twisted or not, since the final particles can be studied with traditional detectors.

Next, since the initial state particles are not anymore momentum eigenstates,
the total momentum of the final system $\bK$ is not fixed.
As a result, the cross section displays certain $\bK$-distribution.
Although it is valid for any wave packet, it becomes most interesting for twisted particles,
as this distribution may contain spectacular interference fringes \cite{Ivanov:2016oue}.
A new dimension in the final-state angular distribution opens up and contains additional information,
not accessible in the plane wave scattering,
which can shine new light on the structure and interactions of hadrons. 

Third, from the theoretical point of view, 
the main calculational difficulty lies not in evaluating Feynman diagrams --- they are the same
as for plane waves --- but in writing the invariant amplitude 
in the general kinematics for initial and final states and then performing the integrations,
especially when polarization vectors and spinors are involved.
To keep the analytic calculations as simple and transparent as possible,
one can first work with Bessel twisted states.
Although they are non-normalizable, one can calculate the differential cross section along the same lines
as for plane waves.
Namely, one defines a finite normalization volume $V$, computes the scattering matrix element,
squares it, regularizes the squares of delta-functions, and obtains the event rate $d\nu$.
Although splitting of $d\nu$ into flux and cross section is not uniquely defined for non-plane-wave collisions \cite{Kotkin:1992bj},
different options have been explored for twisted particles, \cite{Jentschura:2010ap,Jentschura:2011ih,Ivanov:2011kk,Ivanov:2016oue};
in the paraxial approximation, their difference is negligible.
For reader's convenience, we outlined this procedure in the appendix.

Fourth, the drawback of pure Bessel state scattering is that the differential cross section
exhibits non-integrable end-point singularities \cite{Ivanov:2011kk}.
If one uses, instead, transversely normalizable twisted state such as Eq.~\eqref{WP},
the cross section becomes finite, see an explicit comparison in \cite{Ivanov:2016oue}.
The price to pay is that the numerical value of the cross section cannot be safely predicted
as it strongly depends on the details of the colliding particle wave functions.
Thus, the figure of merit for twisted particle scattering is not the absolute value of the cross section 
but the {\em patterns} it exhibits in the final momentum distribution and in its dependence on the total collision energy.
It is worth stressing this important message again:
\begin{itemize}
\item[]
In twisted particle scattering, do not pay attention to the absolute value of the cross section, 
as it is sensitive to the initial state preparation. Look at the kinematic distributions, as they provide
information complementary to the plane wave scattering.
\end{itemize}

\section{Resonance production by scalar twisted particles}
\label{sec:2to1.scalar.case}

\subsection{Pure Bessel beams: general expressions}

We begin our analysis of the $2\to 1$ process by recalling how it is calculated
in the usual plane wave case. 
If the energies and momenta are $(E_i, \bk_i)$ for the two initial particles  
and $(E_K, \bK)$ for the produced particle, the plane wave $S$-matrix amplitude has the form
\be
S_{PW} = i(2\pi)^4\delta(E_1+E_2-E_K) \delta^{(3)}(\bk_1 + \bk_2 - \bK) {{\cal M}(k_1,k_2;K) \over \sqrt{8 E_1 E_2 E_K}}\,.
\label{SPW}
\ee
Here, ${\cal M}(k_1,k_2;K)$ is the plane-wave invariant amplitude 
calculated according to the standard Feynman rules.
Squaring this amplitude, regularizing the squares of delta-functions, and diving by flux, 
as described, for instance, in \cite{LL4} and in the appendix, 
one gets the cross section 
\bea
d\sigma &=& {\pi \delta(E_1+E_2-E_K) \over 4 E_1 E_2 E_K v} |{\cal M}|^2  \, 
\delta^{(3)}(\bk_1 + \bk_2 - \bK) \, d^3 K\,,\nonumber\\[2mm]
\sigma &=& {\pi \delta(E_1+E_2-E_K) \over 4 E_1 E_2 E_K v} |{\cal M}|^2\,.\label{sigma-PW-0}
\eea
Notice the well known features of this cross section:
the final momentum is fixed at $\bK = \bk_1 + \bk_2$ and the dependence
on the total collision energy is via $\delta(E_1+E_2-E_K)$.
The production process occurs only when the initial particles are directly ``at the resonance''.

\begin{figure}[!htb]
\centering
\includegraphics[width=0.5\textwidth]{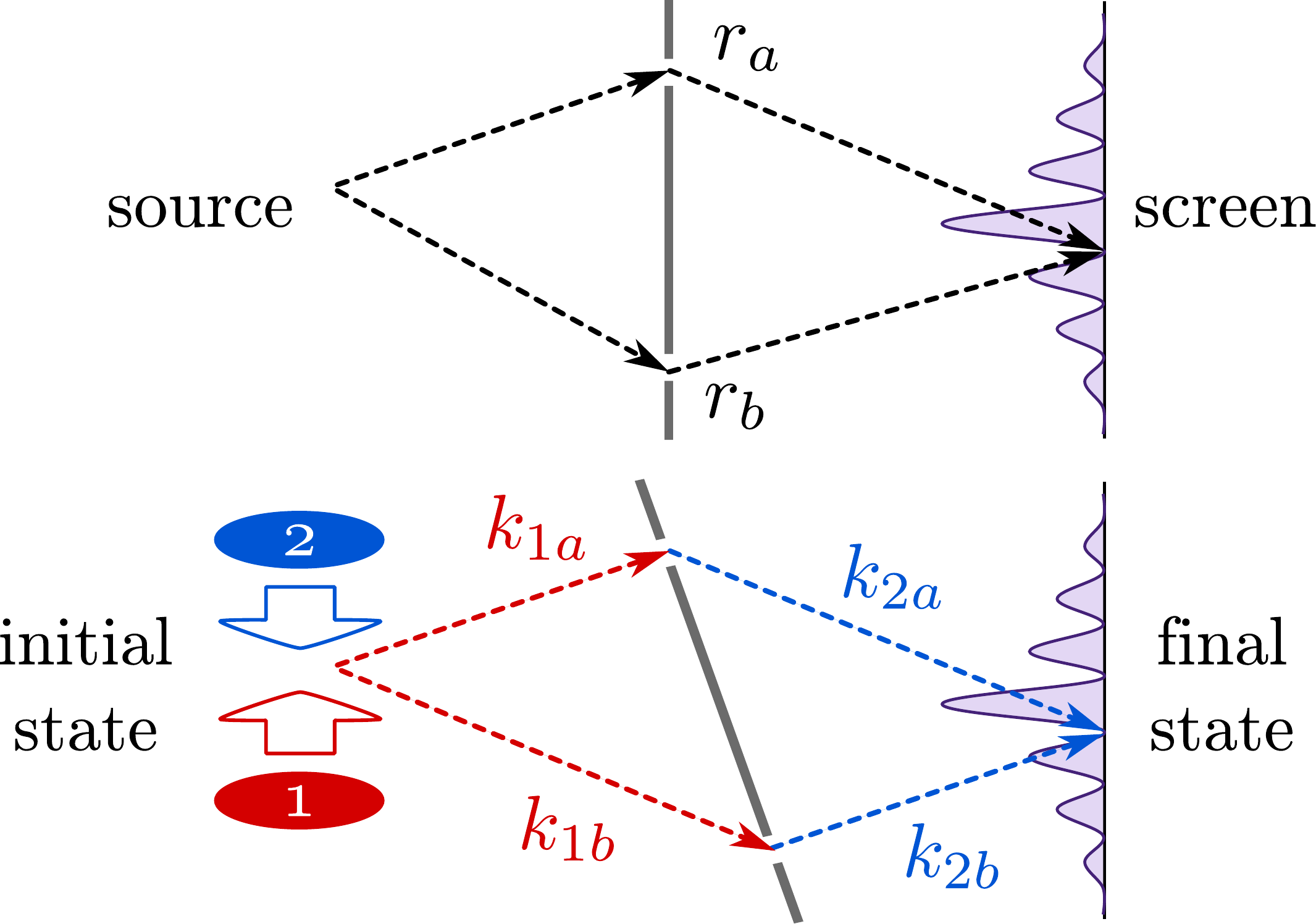}
{\caption{\label{fig-two-slit} Schematic illustration of the classic Young's experiment in coordinate space (upper image)
and of the double-slit experiment in momentum space (lower image), \cite{Ivanov:2016jzt}. In the latter case, 
the arrows show that, in the collision of two Bessel states, only two momentum combinations lead to any final plane-wave state.}}
\end{figure}

Let us now consider collision of two Bessel states $|E_1,\varkappa_1,m_1\rangle$
and $|E_2,\varkappa_2,m_2\rangle$ of spinless and massless particles 
which are defined in the same reference frame and with respect to the same axis $z$.
The final particle with mass $M$ is still described in the basis of plane waves,
and its momentum $\bK$ and energy $E_K$ satisfy $E_K^2 = M^2 + \bK^2$.
We follow the procedure of \cite{Jentschura:2010ap,Ivanov:2011kk,Ivanov:2016jzt},
which adapts the general theory of scattering of non-monochromatic, arbitrarily shaped,
partially coherent beams developed in \cite{Kotkin:1992bj} to the collisions of Bessel twisted states,
see also the appendix.
The $S$-matrix element of this process is 
\be
S = \int {d^2 \bk_{1\perp} \over (2\pi)^2} {d^2 \bk_{2\perp} \over (2\pi)^2} 
a_{\varkappa_1 m_1}(\bk_{1\perp}) a_{\varkappa_2, -m_2}(\bk_{2\perp}) S_{PW}\,.\label{S-tw}
\ee
The negative sign in front of $m_2$ reflects the fact that the second particle propagates on average
in the $-z$ direction.
Substituting in \eqref{S-tw} the Fourier amplitudes of the Bessel states,
we get
\be
S = i (2\pi)^4 \fr{\delta(\Sigma E) \delta(\Sigma k_z)}{\sqrt{8 E_1 E_2 E_K}} 
{(-i)^{m_1-m_2} \over (2\pi)^3\sqrt{\varkappa_1\varkappa_2}} \cdot {\cal J}\,,\label{S-tw2}
\ee
where $\delta(\Sigma E) \equiv \delta(E_1+E_2-E_K)$, $\delta(\Sigma k_z) \equiv \delta(k_{1z}+k_{2z}-K_z)$.
The twisted amplitude ${\cal J}$ is defined as
\bea
{\cal J} &=& \int d^2 \bk_{1\perp} d^2 \bk_{2\perp} \, e^{im_1\varphi_1 - im_2\varphi_2}\,
\delta(|\bk_{1\perp}|-\varkappa_1) \delta(|\bk_{2\perp}|-\varkappa_2)
\delta^{(2)}(\bk_{1\perp}+\bk_{2\perp} - \bK_\perp)\cdot {\cal M}\nonumber\\
&=&\varkappa_1\varkappa_2 \int d\varphi_1 d\varphi_2\, e^{im_1\varphi_1 - im_2\varphi_2}\,
\delta^{(2)}(\bk_{1\perp}+\bk_{2\perp} - \bK_\perp)\cdot {\cal M}\,,\label{J}
\eea
and it has the same dimension as the matrix element ${\cal M}$.
\begin{figure}[ht]
\centering
\includegraphics[width=0.75\textwidth]{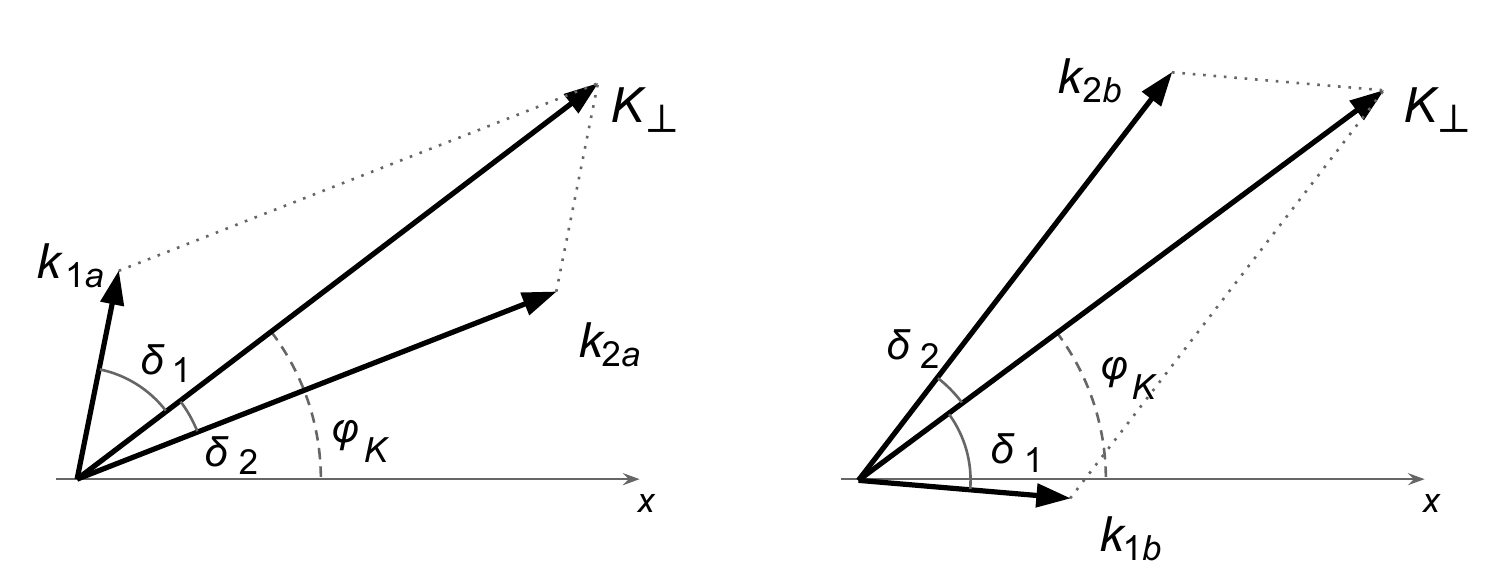}
{\caption{\label{fig-2configurations} The two kinematic configurations in the transverse plane which satisfy
the transverse momentum conservation law in collision of two Bessel states.}}
\end{figure}
Since it contains an equal number of integrations and delta-functions, it can be calculated exactly \cite{Ivanov:2011kk}. 
It is non-zero only if the moduli of the transverse momenta $\varkappa_i \equiv |\bk_{i\perp}|$ 
and $K \equiv |\bK_\perp|$ satisfy the triangle inequalities
\be
|\varkappa_1 - \varkappa_2| \le K \le \varkappa_1 + \varkappa_2\,.\label{ring}
\ee
They form a triangle with the area
\be
\Delta 
= {1 \over 4} \sqrt{2 K^2\varkappa_1^2 + 2 K^2\varkappa_2^2 + 2\varkappa_1^2\varkappa_2^2 
- K^4 - \varkappa_1^4 - \varkappa_2^4}\,.\label{area}
 \ee
Out of many plane wave components ``stored'' in the initial twisted particles, 
the integral \eqref{J} receives contributions from exactly two plane wave combinations shown in Fig.~\ref{fig-2configurations}
with the following azimuthal angles:
\bea
&& \mbox{configuration a:} \quad \varphi_1 = \varphi_{K} + \delta_1\,,\quad \varphi_2 = \varphi_{K} - \delta_2\,,\nonumber\\
&& \mbox{configuration b:} \quad \varphi_1 = \varphi_{K} - \delta_1\,,\quad \varphi_2 = \varphi_{K} + \delta_2\,.\label{phi12}
\eea
Notice that
\be
\delta_1 = \arccos\left({\varkappa_1^2 + K^{2} - \varkappa_2^2 \over 2\varkappa_1 K}\right)\,,
\quad
\delta_2 = \arccos\left({\varkappa_2^2 + K^{2} - \varkappa_1^2 \over 2\varkappa_2 K}\right)
\label{delta_i}
\ee
are the inner angles of the triangle with the sides $\varkappa_1$, $\varkappa_2$, $K$;
they are not azimuthal variables. 
The result for the twisted amplitude ${\cal J}$ can then be compactly written as
\be
{\cal J} = e^{i(m_1 - m_2)\varphi_{K}}{\varkappa_1 \varkappa_2 \over 2\Delta}
\left[{\cal M}_{a}\, e^{i (m_1 \delta_1 + m_2 \delta_2)} + {\cal M}_{b}\, e^{-i (m_1 \delta_1 + m_2 \delta_2)}\right]\,.\label{J2}
\ee
Notice that the plane-wave amplitudes ${\cal M}_{a}$ and ${\cal M}_{b}$ are calculated for the two distinct initial momentum 
configurations shown in Fig.~\ref{fig-2configurations} but for the same final momentum $\bK$.
They exhibit two distinct paths in momentum space to arrive at the same final state from the initial twisted states.
In a sense, scattering of twisted Bessel states represents the momentum space analog of the Young double-slit
experiment as illustrated in Fig.~\ref{fig-two-slit}.  

Since in this paper we focus on kinematic features of twisted particle annihilation,
we take the simplest example of pointlike interaction among the three scalar particles. 
In this case, the invariant amplitude ${\cal M} = g$ does not depend on the momenta,
and the twisted amplitude ${\cal J}$ in \eqref{J2} simplifies further:
\be
{\cal J} = e^{i(m_1 - m_2)\varphi_{K}} g {\varkappa_1 \varkappa_2 \over \Delta}\cos(m_1 \delta_1 + m_2 \delta_2)\,.
\label{J-scalar}
\ee

\subsection{Pure Bessel beams: kinematic distributions}

Let us repeat the expression for the $S$-matrix amplitude in the pure Bessel beam case as
\be
S \propto \delta(E_1+E_2-E_K)\, \delta(k_{1z}+k_{2z}-K_z) \cdot {\cal J}\,,\label{S-tw-again}
\ee
where ${\cal J}$ is given by \eqref{J} or \eqref{J-scalar}.
From now on, we deliberately omit the prefactors to stress, as discussed in the section~\ref{subsection:broad-picture},
that the figure of merit is not the absolute value of the cross section but its kinematic distribution.
The exact expressions, if needed, can be found in \cite{Ivanov:2011kk,Ivanov:2016oue}. 
Squaring \eqref{S-tw-again} and performing the standard regularization of the squares of the two delta-functions \cite{LL4},
we obtain the (generalized) cross section in the form
\begin{eqnarray}
d\sigma &\propto& \delta(E_1+E_2-E_K)\, \delta(k_{1z}+k_{2z}-K_z)\, |{\cal J}|^2 \, d^3K \nonumber\\ 
&=& \delta(E_1+E_2-E_K)\, |{\cal J}|^2 \, d^2\bK_\perp\,.\label{dsigma-tw}
\end{eqnarray}
Unlike the plane-wave $2\to 1$ collision cross section \eqref{sigma-PW-0}, 
where the momentum of the final particle is completely fixed, here we have 
a distribution over $\bK_\perp$. 

If we study production of a final particle with mass $M$, 
the value of $K\equiv |\bK_\perp|$ is not a free variable but is fixed by the energy conservation:
\be
E_K^2 = (E_1+E_2)^2 = K^2 + (k_{1z}+k_{2z})^2 + M^2\,.
\ee
This fixes the polar angle of the produced resonance:
\begin{equation}
\cos\theta_K = \fr{K_z}{\sqrt{(E_1+E_2)^2-M^2}}\,.\label{theta-K}
\end{equation}
The cross section reduces to 
\be
d\sigma \propto {1 \over \Delta^2} \cos^2(m_1 \delta_1 + m_2 \delta_2) \,d\varphi_K\label{dsigma-tw3}
\ee
with a uniform $\varphi_K$ distribution.

\begin{figure}[!htb]
	\centering
	\includegraphics[height=5cm]{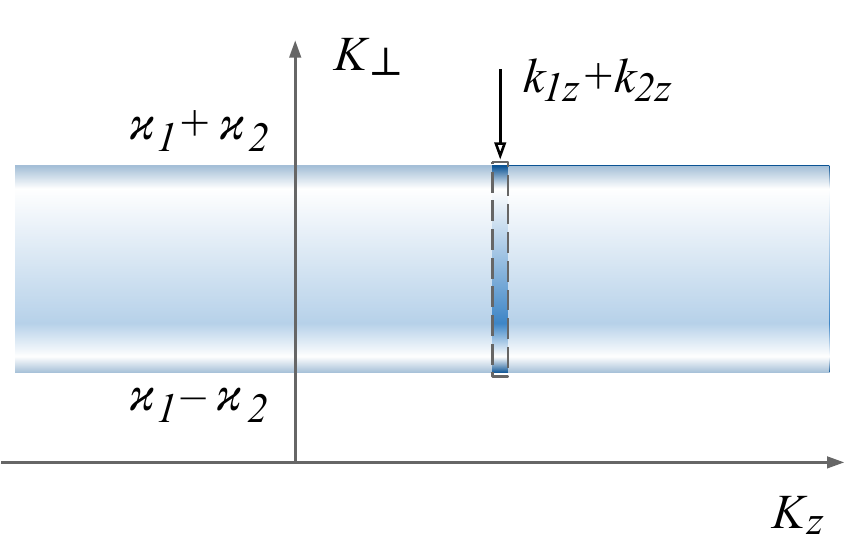}\hspace{1cm}
	\includegraphics[height=5cm]{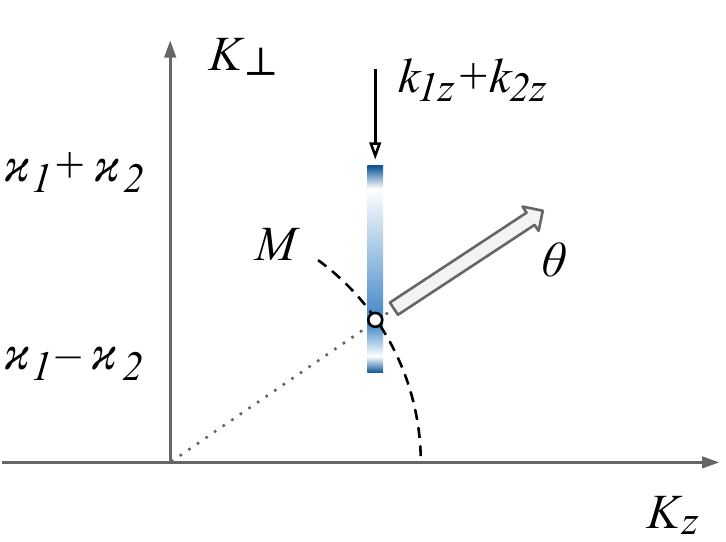}
	\caption{Final momenta available for production of a resonance of mass $M$ in collision of two twisted Bessel states.
	{\em Left}: the pale band shows $|{\cal J}|^2$, 
	while the arrow indicates the longitudinal momentum fixed by $\delta(k_{1z}+k_{2z}-K_z)$.
	{\em Right}: the energy conservation restricts the three-momentum of the produced particle with mass $M$ to lie on the dashed arc;
	the point where the arc crosses the kinematically available interval defines the produced resonance polar angle.
}
	\label{fig:2d-illustration}
\end{figure}

Fig.~\ref{fig:2d-illustration} illustrates the above description in the $(K, K_z)$ space.
The pale blue band in the left image shows $|{\cal J}|^2$, which is defined in the $K$ range \eqref{ring}
and which displays the interference fringes and the end-point singularities.
The value of $|{\cal J}|^2$ does not depend on the longitudinal momentum $K_z$.
The value of $K_z$ is constrained, instead, by the longitudinal delta-function $\delta(k_{1z}+k_{2z}-K_z)$
in the cross section. Thus, the final momentum space available for resonance production
in collision of two initial state monochromatic Bessel particles is restricted to the (infinitesimally) narrow 
band schematically indicated in the left figure by the arrow pointing to the dashed window.

When a final particle with mass $M$ is produced, the energy conservation fixes the absolute value 
of its three-momentum $|\bK| = \sqrt{K^2 + K_z^2}$, which is indicated by the dashed arc in the right plot of Fig.~\ref{fig:2d-illustration}.
If the total energy lies between
\begin{equation}
\sqrt{(\varkappa_1 - \varkappa_2)^2 + M^2+K_z^2} \le E_1+E_2 \le \sqrt{(\varkappa_1 + \varkappa_2)^2 + M^2+K_z^2}\,,\label{E-range}
\end{equation}
the arc crosses the available momentum range and the particle with mass $M$
can indeed be produced. Since its $K_z$ and $K$ are fixed, it is emitted at the fixed polar angle given by \eqref{theta-K}.

Let us now imagine that one can continuously change the initial particle kinematics
and perform a scan over the total collision energy $E_1+E_2$.
Unlike the plane-wave case, where the cross section is non-zero only directly at the resonance, 
here one observes a non-trivial distribution across the entire energy interval \eqref{E-range}. 
As one changes the total energy, one sees that the value of the cross section goes up and down, 
as one slides across the interference fringes
given by $\cos^2(m_1 \delta_1 + m_2 \delta_2)$, or reaches the end-point enhancement due to $1/\Delta^2$. 
The number and the heights of the fringes
depend on the OAM values $m_i$ and on the absolute values of the transverse momenta $\varkappa_i$ 
of the initial twisted particles.
We stress that all these fringes are exhibited by a {\em single} resonance, whose production
becomes stronger or weaker as one scans the total energy.

\begin{figure}[!htb]
	\centering
	\includegraphics[width=0.32\textwidth]{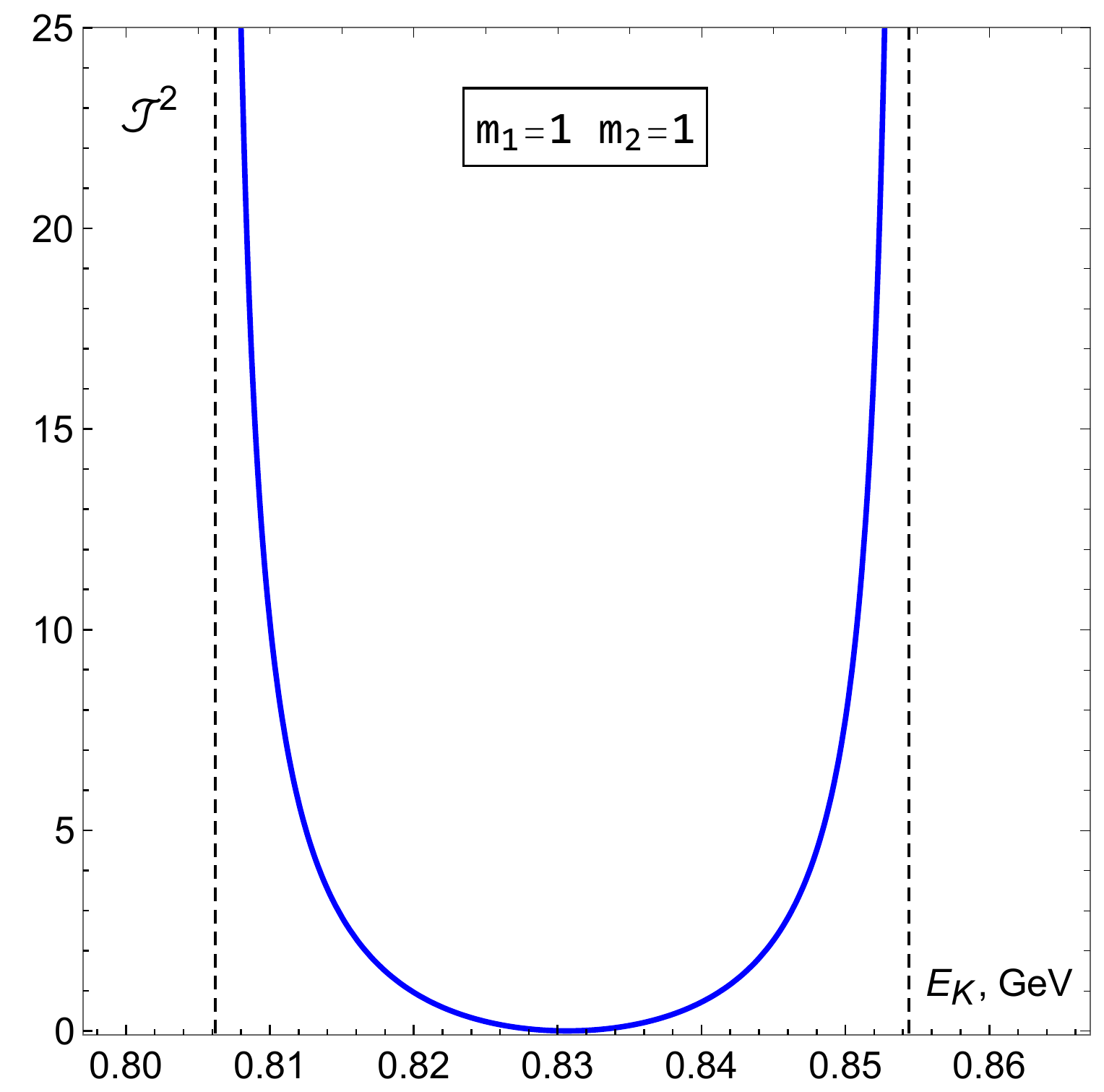}
	\includegraphics[width=0.32\textwidth]{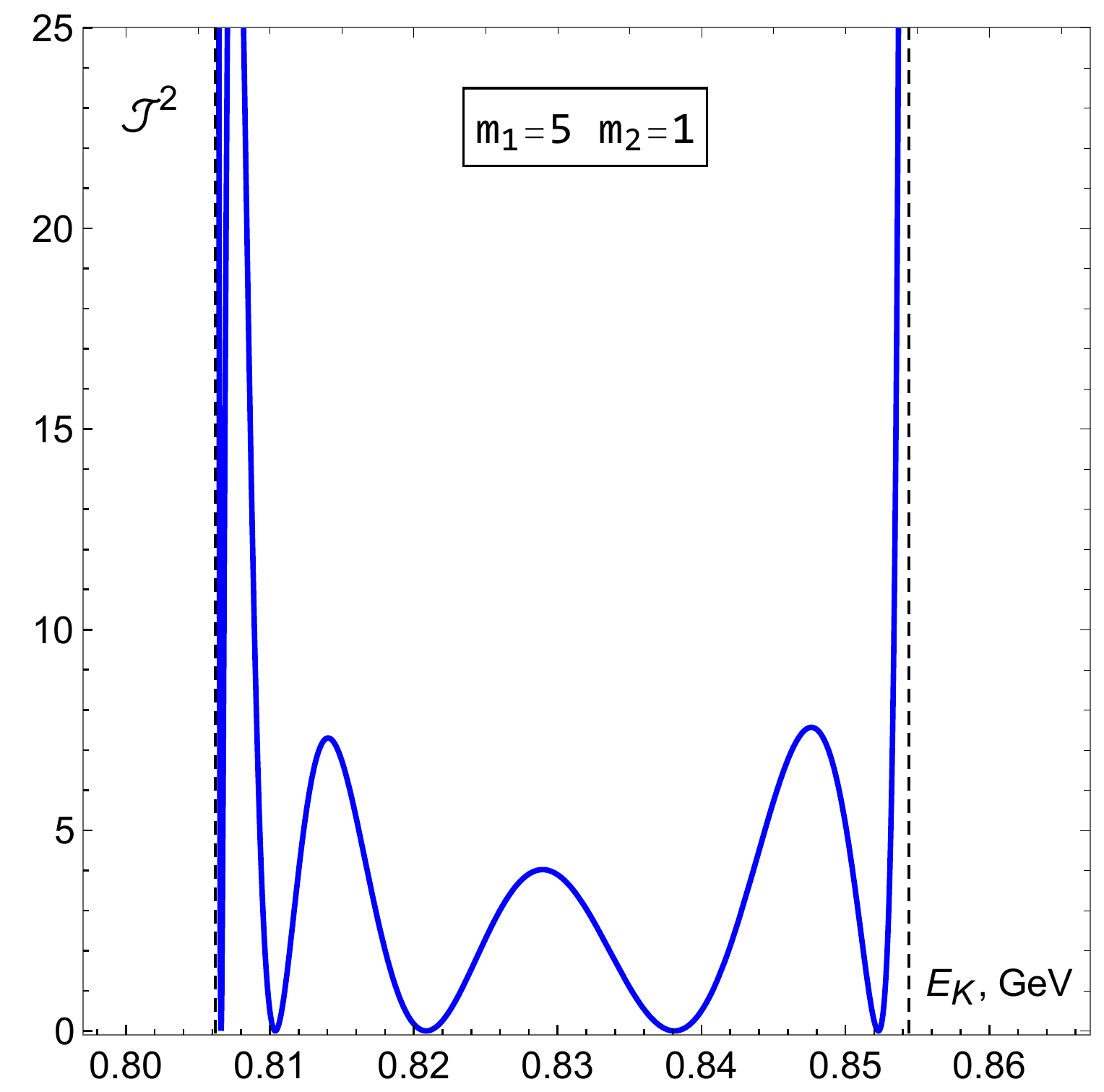}
	\includegraphics[width=0.32\textwidth]{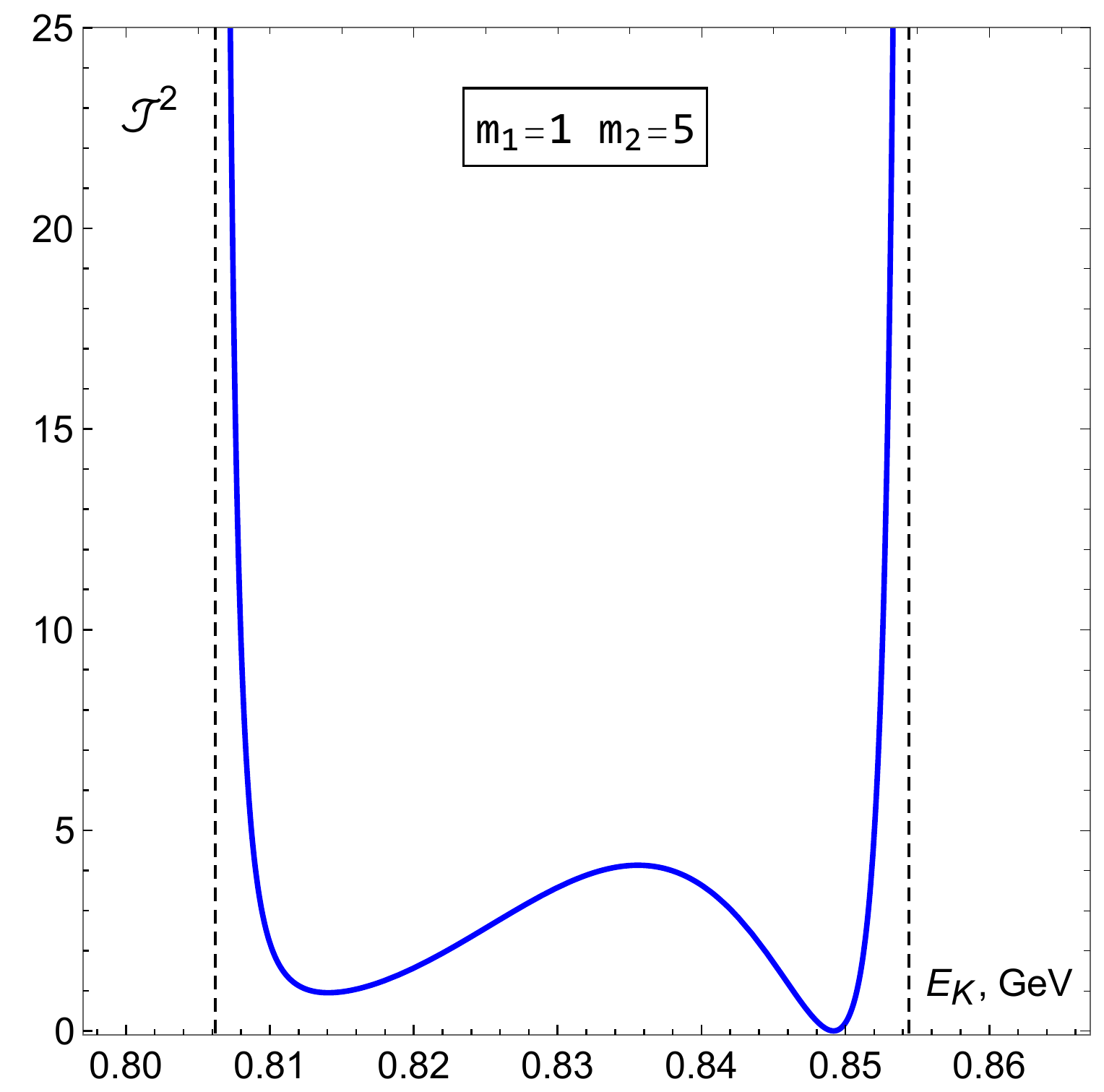}
	\caption{
		The resonance production cross section, 
		which is proportional to $|{\cal J}|^2$ defined in \eqref{J-scalar} and is given in arbitrary units, as a function
		of the total energy of the colliding scalar particles for the three kinematic configurations
		listed in \eqref{3-scalar-examples-1} and \eqref{3-scalar-examples-2}.
	}
	\label{fig-scalar-unsmeared}
\end{figure}

As an example, we show in Fig.~\ref{fig-scalar-unsmeared} this distribution for 
production of a resonance with mass $M= 0.8\, \GeV$ in collision of two twisted massless
particles with the following parameters
\be
\varkappa_1 = 0.1\, \GeV\,, \quad \varkappa_2 = 0.2\, \GeV\,, \quad k_{1z} = - k_{2z}\,,\label{3-scalar-examples-1}
\ee
and for three choices of $m_i$:
\be
(m_1, m_2) = (1,\, 1)\,,\quad (5,\, 1)\,,\quad (1,\, 5)\,.\label{3-scalar-examples-2}
\ee 
The fringes are especially visible for the case when $m$ corresponding to the {\em smaller} $\varkappa$
is large. This is to be expected. When one scans over the total energy, the transverse momentum $K = |\bK_\perp|$ 
corresponding to the same mass $M$ of the produced particle varies from the minimal
to the maximal allowed values for given $\varkappa_1$ and $\varkappa_2$. The inner angles of the triangle
$\delta_1$ and $\delta_2$ defined in \eqref{delta_i} behave differently: the angle adjacent to the smaller
$\varkappa$ sweeps from 0 to $\pi$, while the other inner angle only reaches certain maximal value
and then decreases to zero.

It is appropriate to mention here an ambiguity in how we perform the scan over the total energy $E_1+E_2$.
In order to indicate how the total energy is distributed between $E_1$ and $E_2$,
we need to choose a ''scan trajectory'' on the $(E_1, E_2)$ plane, 
which can be done by fixing a suitable kinematic variable. 
In the numerical example just shown we relied on fixed $\varkappa_i$ and on the auxiliary relation $k_{1z} + k_{2z} = 0$
which corresponds to the fixed production polar angle $\theta_K = \pi/2$.
Then, for each value of $E_1+E_2$, the individual energies $E_1$ and $E_2$ can be computed, defining the scan trajectory.
However this is only one of the many possible choices. 
One could alternatively fix $E_1 = E_2$, or freeze $E_1$ and vary only $E_2$.
All these options lead to plots of the cross section which will be all similar but their details 
such as the positions and the heights of the fringes may differ.
There is no uniquely preferred option; ultimately, the exact scan trajectory will depend
on what is feasible experimentally.

\subsection{Realistic twisted beams}

The expression for the cross section \eqref{dsigma-tw} and the exact evaluation of ${\cal J}$ in \eqref{J2}
were obtained for pure Bessel states. As explained above,
these states are not normalizable and lead to the $1/\Delta^2$ singularity in the cross section, 
which diverges if integrated up to the end points.

For realistic twisted states normalizable in the transverse plane
this singularity disappears. One possible choice of such a realistic state
is a monochromatic $\varkappa$-smeared wave packet given in Eq.~\eqref{WP}.
One does not need to repeat the derivation of the cross section; one can just apply this smearing procedure
with the functions $f_1(\varkappa_1)$ and $f_2(\varkappa_2)$ to $S$-matrix amplitude \eqref{S-tw2}.
Therefore, instead of ${\cal J} \cdot \delta(k_{1z}+k_{2z}-K_z)$ for pure Bessel states, 
we now need to evaluate its smeared counterpart which we define as:
\begin{equation}
\lr{{\cal J}} = \int\limits_0^{E_1} d\varkappa_1 \int\limits_0^{E_2} d\varkappa_2 f_1(\varkappa_1) f_2(\varkappa_2) \delta(k_{1z} + k_{2z} - K_z) 
{{\cal J}(\varkappa_1,\varkappa_2) \over \sqrt{\varkappa_1\varkappa_2}}\,.\label{J-smeared}
\end{equation}
In numerical calculations, we use the Gaussian smearing functions of the following form:
\begin{eqnarray}
f_i(\varkappa_i)=n_{i}
\sqrt{\varkappa_i}
\exp\left[
-\frac{(\varkappa_i-\bar\varkappa_i)^2}{2\si_i^2}\right]\,.
\end{eqnarray}
The normalization condition $\int_0^{E_i} d\varkappa |f_i(\varkappa)|^2=1$
fixes the normalization constants $n_i$, and for $E_i\gg \bar\varkappa_i\gg \si_i$, they are approximately equal to
$n_i=1/\sqrt{\sqrt{\pi}\si_i\bar\varkappa_i}$.
The Bessel state limit is restored when $\sigma_i \to 0$:
\begin{eqnarray}\label{eq:weak.limit}
\sqrt{\frac{\bar\varkappa_1\bar\varkappa_2}{4\pi\si_1\si_2}}
\lr{{\cal J}}  \mathop{\rightarrow}\limits_{\si_i \to 0} 
\de(k_{1z}+k_{2z}-K_z)\,\cdot {\cal J}\,.
\end{eqnarray}
The longitudinal delta-function in \eqref{J-smeared} can be used to remove the $\varkappa_2$ integration:
\begin{equation}
\lr{{\cal J}} = 
\int \!d\varkappa_1 
f_1(\varkappa_1) f_2(X_2)
\frac{{\cal J}(\varkappa_1,X_2)}{\sqrt{\varkappa_1}X_2^{3/2}} 
\sqrt{E_2^2-X_2^2}\,, \label{J-smeared-2}
\end{equation}
where $\varkappa_2$ is replaced by $X_2$ which depends on $\varkappa_1$ and $K_z$:
\begin{equation}
X_2 = \left[E_2^2-\left(\sqrt{E_1^2-\varkappa_1^2}-K_z\right)^2\right]^{1/2}.\label{X2}
\end{equation}
It is assumed that the integration limits on $\varkappa_1$ are such that 
both square roots in \eqref{X2} are well defined
and the condition \eqref{ring} is fulfilled.

Since the longitudinal momentum delta-function disappears, the cross section can be written as 
\begin{equation}
d\sigma \propto |\lr{{\cal J}}|^2 \delta(E_1+E_2-E_K) \, d^3K\,.\label{dsigma-tw-2}
\end{equation}
Removing the energy delta-function, one can obtain either a non-trivial angular distribution over a finite range
of polar angles,
\begin{equation}
d\sigma \propto E_K^2 \beta_K\, |\lr{{\cal J}}|^2\, d\Omega_K\,,\label{dsigma-tw-3}
\end{equation}
or a distribution over transverse momenta with the correlated change of $K_z = \sqrt{E_K^2 - K^2 - M^2}$:
\begin{equation}
d\sigma \propto {E_K \over K_z}\, |\lr{{\cal J}}|^2\, d^2 \bK_\perp\,.\label{dsigma-tw-4}
\end{equation}

\begin{figure}[!htb]
	\centering
	\includegraphics[height=5.5cm]{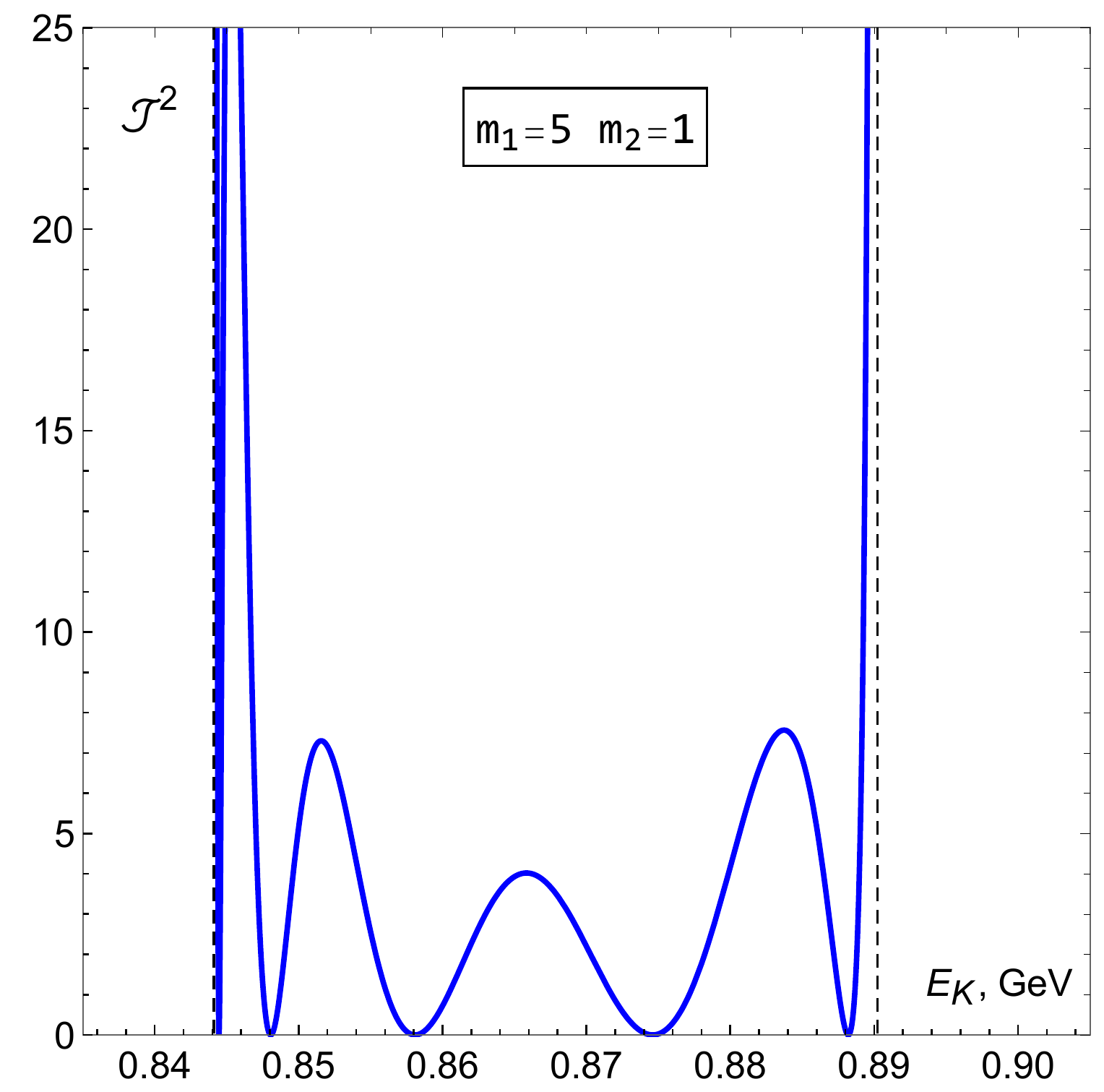}
	\includegraphics[height=5.5cm]{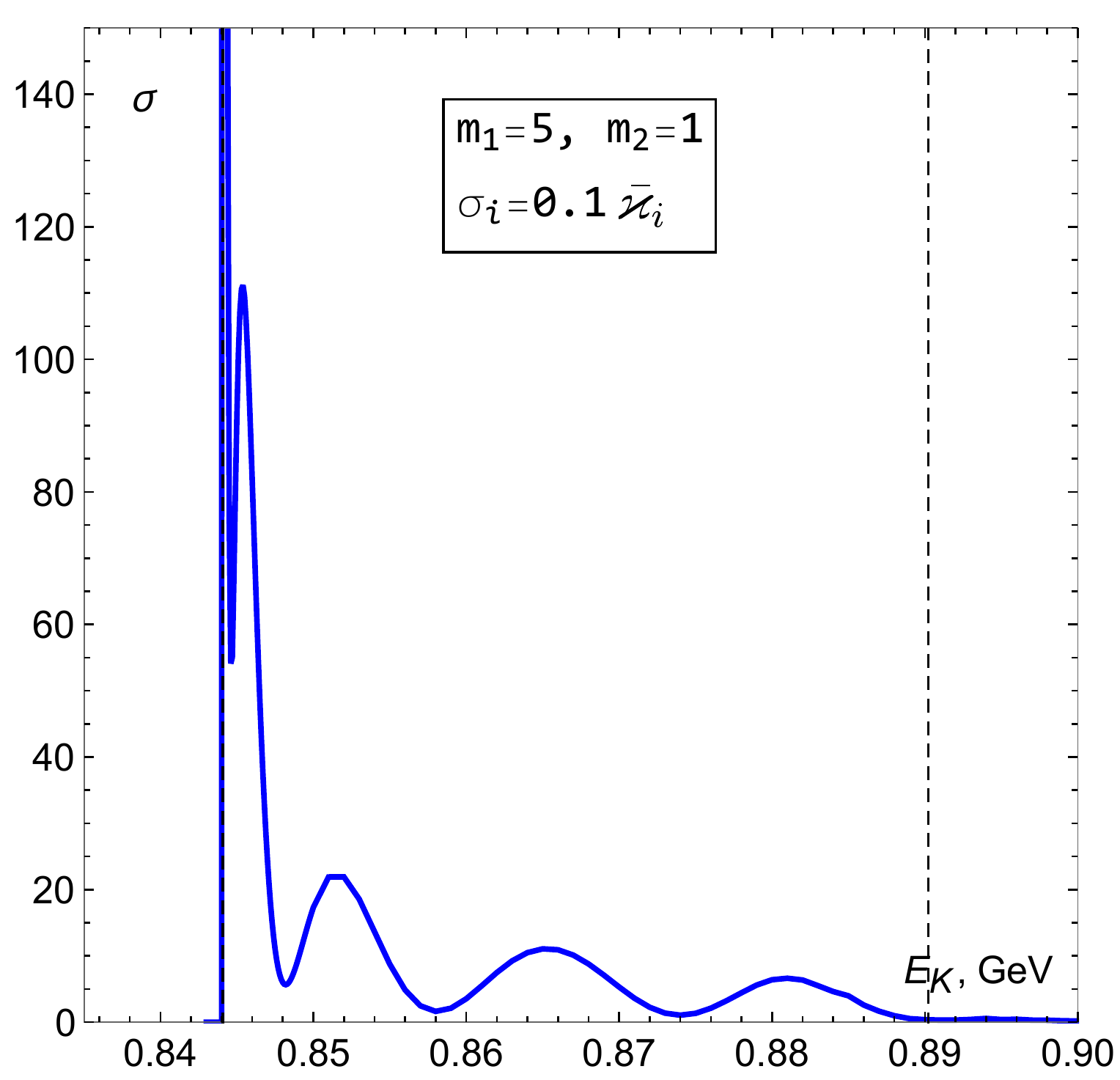}
	\includegraphics[height=5.5cm]{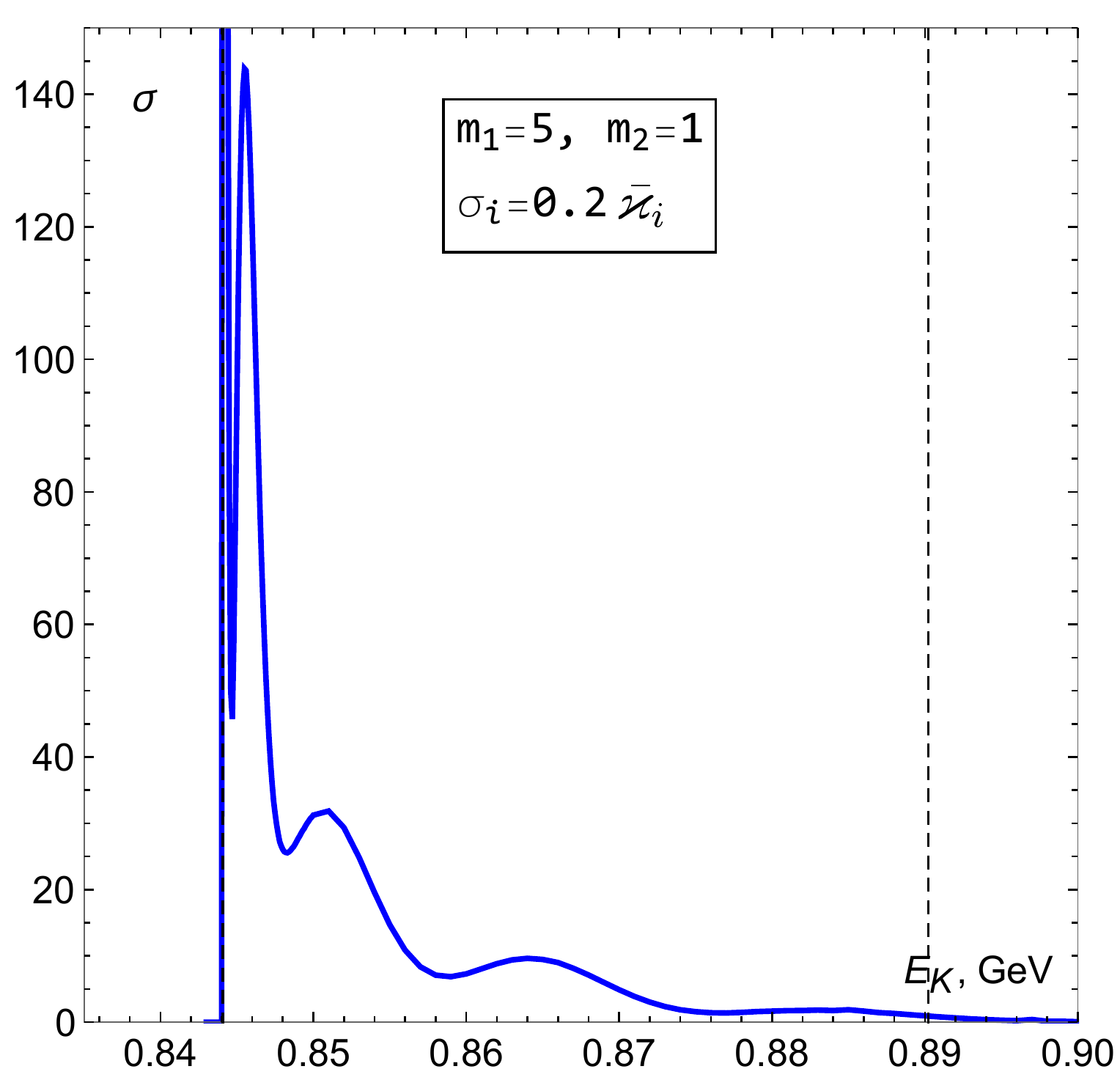}
	\caption{The cross section as a function of the total collision energy for the kinematic parameters \eqref{smearing-option-1}  
		with three values of Gaussian smearing: $\sigma_i/\bar\varkappa_{i} = 0$ (left), $0.1$ (middle), and $0.2$ (right). 
		The scan trajectory is fixed by the relation $\bar{K_z} = -0.25$ GeV.
		The cross sections are represented, in arbitrary units, by $|{\cal J}|^2$ in \eqref{J-scalar} for the left plot,
		and $\int |\lr{{\cal J}}|^2 d\cos\theta_K $ defined in \eqref{J-smeared-2} for the middle and right plots.
	}
	\label{fig-scalar-smeared}
\end{figure}

	To demonstrate the effect of smearing on the visibility of the interference fringes in $\sigma(E_K)$, 
we present in Fig.~\ref{fig-scalar-smeared} the cross section as a function of energy 
calculated with the following kinematic parameters: 
\begin{equation}
\label{smearing-option-1}
(m_1, m_2) = (5, 1)\,, \quad \bar{\varkappa}_{1}=0.1~\GeV, \quad  \bar{\varkappa}_{2}=0.2~\GeV, \  
\end{equation}
and with different amount of smearing: $\sigma_i/\bar\varkappa_{i} = 0$, $0.1$, and $0.2$
for the left, middle, and right plots, respectively.
In this case, we selected the scan trajectory by the condition that $\bar K_z  = -0.25$ GeV
is fixed during energy scan. 
Here, $\bar K_z = \bar k_{1z} + \bar k_{2z}$, where $\bar k_{iz} = \sqrt{E_i^2 - \bar\varkappa_i^2}$. 
Notice that the left plot in Fig.~\ref{fig-scalar-smeared}, which corresponds to the non-smeared case, 
is very similar to Fig.~\ref{fig-scalar-unsmeared}, middle plot,
as they differ only by the choices of $\bar{K_z}$.
One sees that the interference fringes remain well visible for a 20\% smearing.

The plots in Fig.~\ref{fig-scalar-smeared} display another striking phenomenon:
smearing over $\varkappa_i$ strongly blurs the cross section around the upper energy boundary of Eq.~\eqref{E-range}
but keeps the lower energy boundary as sharp as before.
At first, this behavior seems counter-intuitive. 
To give a qualitative explanation of what is happening at the lower boundary,
let us fix a particular value $K_z$ and vary $\varkappa_i$.
A fixed $K_z = k_{1z} + k_{2z} = k_{1z} - |k_{2z}|$ implies that $\varkappa_1$ and $\varkappa_2$
must co-vary, that is, they either both grow or both decrease.
This leads to a significant variation of $\varkappa_1 + \varkappa_2$ but to a much more reduced 
variation of $|\varkappa_1 - \varkappa_2|$.
As a result, the upper energy limit is expected to blur while the lower one is more stable.

Quantitatively, let us see how the lower energy boundary $E_{min}$ varies with $\varkappa_i$ at fixed $E_i$.
\begin{equation}
d (E^2_{min}/2) = K_z (d k_{1z} + d k_{2z}) + (\varkappa_1 - \varkappa_2) (d\varkappa_1 - d\varkappa_2)\,.
\end{equation}
Since $E_1$ and $E_2$ are fixed, we get
\begin{equation}
k_{1z} dk_{1z} + \varkappa_1 d\varkappa_1 = 0\,, \quad
k_{2z} dk_{2z} + \varkappa_2 d\varkappa_2 = 0\,.
\end{equation}
As a result, 
\begin{equation}
d (E^2_{min}/2) = - (k_{1z} \varkappa_2 + k_{2z} \varkappa_1) \left(\frac{d\varkappa_1}{k_{1z}} + \frac{d\varkappa_2}{k_{2z}}\right)\,.
\end{equation}
Since $k_{2z} < 0$, the first bracket becomes zero when $\varkappa_1/k_{1z} = \varkappa_2/|k_{2z}|$,
that is, when the opening angles of the two twisted particle cones coincide.
When fixing the scan trajectory by $\bar K_z = -0.25$ GeV and using parameters \eqref{smearing-option-1},
we satisfied this condition to a good accuracy across the relevant range of $E_i$.
This is why the lower energy boundary stays almost invariant even for significant smearing of $\varkappa_i$.
Choosing a significantly different value of $\bar K_z$ leads to a visible smearing of the lower boundary
and to larger smearing of the interference fringes.

\subsection{Two ways to reveal the fringes}

Numerical investigations show that the visibility of fringes in the $\sigma(E_K)$ plot depends on the kinematic arrangements:
in certain cases they are well visible, in other cases they are washed out.
In this subsection we provide a transparent intuitive understanding of this behavior.
In fact, the interference fringes, 
which are the hallmark feature of the two-path interference of the twisted particle collisions,
are present in the 2D plane $(K, K_z)$ in all cases.
One just needs a suitable variable for each choice of the initial parameters to reveal them.

Indeed, switching from the pure Bessel to realistic twisted states reshapes the range of final particle
momenta available. The quantity $|\lr{{\cal J}}|^2$ replacing $|{\cal J}|^2 \cdot \delta(k_{1z}+k_{2z}-K_z)$
now displays not only a $K$ distribution but also a $K_z$ distribution of finite width
peaked around $\bar K_z = \bar k_{1z} + \bar k_{2z}$. 
Moreover, the two distributions are correlated.
This is why the interference fringes must still persist in the $(K,K_z)$ plane
in the form of oblique stripes. The orientation of these stripes
is not fixed and, depending on the kinematic arrangements, can be adjusted. 
Additionally, one can shift the location of these fringe pattern along axis $K_z$
by adjusting the energies of the initial particles, which has a profound effect on visibility 
of the fringes.

\begin{figure}[!htb]
	\centering
	\includegraphics[width=0.8\linewidth]{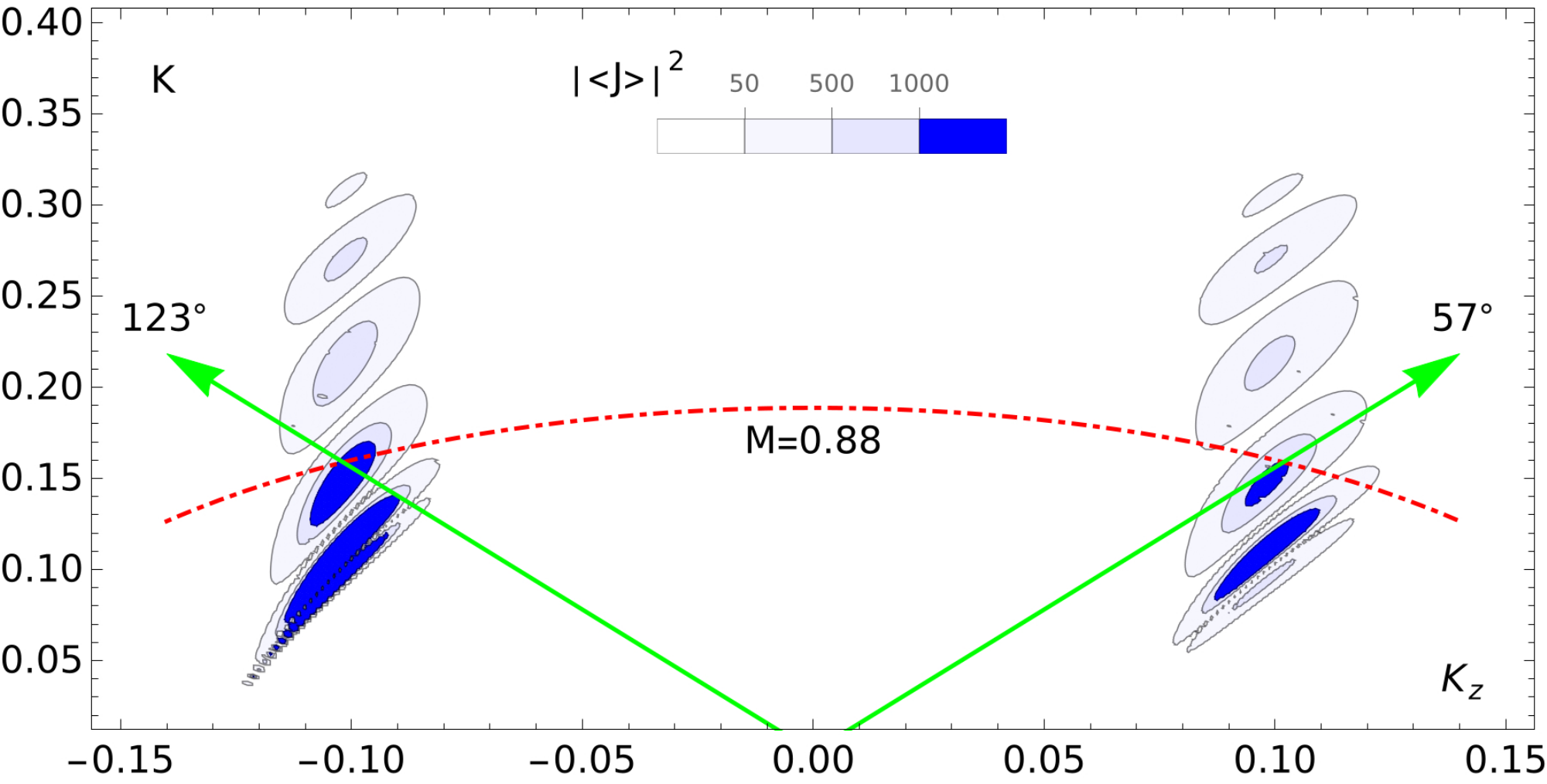}
	\caption{Even for smeared twisted states, the cross section displays clean interference fringes
		on the plane $(K,K_z)$. Shown here is $|\lr{{\cal J}}|^2$ evaluated for the kinematic parameters \eqref{smearing-option-1} 
		with two choices of energies: $E_1 < E_2$ (left part) and $E_1 > E_2$ (right part).
		The constant $\theta_K$ rays (solid line) and the constant $M$ arc (dashed line) are also shown. 
	}
	\label{fig:2d-smeared}
\end{figure}

In Fig.~\ref{fig:2d-smeared} we show two illustrative numerical examples of this phenomenon.
The plot shows the value of $|\lr{{\cal J}}|^2$ evaluated for the kinematic parameters of \eqref{smearing-option-1}
with $\sigma_i/\bar\varkappa_{i} = 0.1$ and with two options for the initial particle energies:
$E_1=0.386$ GeV, $E_2 =0.514$ GeV (left plot, $K_z < 0$) and 
$E_1=0.48$ GeV, $E_2 = 0.42$ GeV (right plot, $K_z > 0$). 
In these two cases, the overall shape of the fringes does not change much,
but their orientation with respect to the origin changes dramatically.
In the example shown in the right half of the plot, the fringes are approximately aligned with the constant $\theta_K$ lines
and, therefore, they are well visible in the angular distribution at fixed energy.
The energy scan of the total cross section $\sigma(E_K)$ will not reveal significant oscillations in this case.
In the example shown in the left half of the plot, the fringes approximately follow the arc corresponding 
to a final particle of fixed mass $M=0.88~\GeV$. 
As a result, the particle is produced in a wide region of polar angles $\theta$ and its production
cross section does not show any particularly strong oscillations in the angular distribution.
However the energy scan of the integrated cross section $\sigma(E_K)$ will show clean peaks, 
as the arc passes through a bright or dark interference fringe.
This is indeed revealed by Fig.~\ref{fig-scalar-smeared}.
Notice also the sharp lower boundary in the left part of the plot but not in the right part.

The net result is that there are two complementary observables capable of revealing the interference fringes
in the twisted particle annihilation: the angular distribution $d\sigma/d\cos\theta_K$ at fixed energies
and the energy dependence of the integrated cross section $\sigma(E_K)$.
According to the choice of the initial $E_i$, $\varkappa_i$ and $m_i$,
either observable can be the most indicative quantity.

\section{Intrinsic mass spectrometry}\label{section-mass-spectrometer}


A unique feature of twisted particle collisions is that the final particle momentum is not fixed.
As a result, it open the door to a phenomenon which is completely impossible in the plane wave $2\to 1$ process:
simultaneous $s$-channel production of two or more resonances with different masses in {\em fixed energy} annihilation.

\begin{figure}[!htb]
	\centering
	\includegraphics[height=6cm]{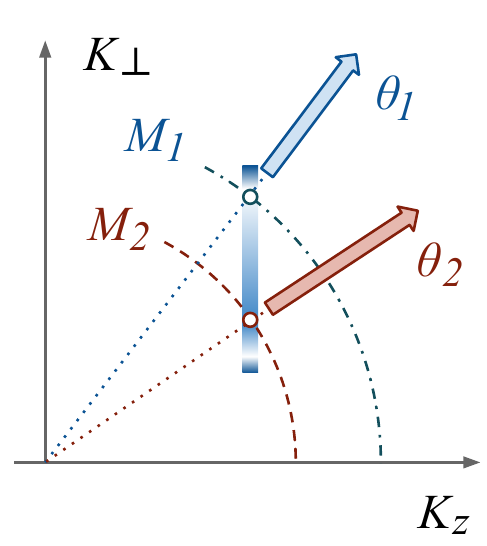}
	\caption{Same as Fig.~\ref{fig:2d-illustration}, right, but for two resonances with masses $M_1$ and $M_2>M_1$.
	}
	\label{fig:2d-M1M2}
\end{figure}

The main idea is illustrated by Fig.~\ref{fig:2d-M1M2}. Let us return, for simplicity,
to the pure Bessel monochromatic beams. Since the total energy is fixed, 
two resonances with masses $M_1$ and $M_2>M_1$ correspond to two different arcs on the $(K, K_z)$ plane.
However, if their masses satisfy
\begin{equation}
\sqrt{(E_1+E_2)^2 - K_z^2 - (\varkappa_1 + \varkappa_2)^2} \le M \le \sqrt{(E_1+E_2)^2 - K_z^2 - (\varkappa_1 - \varkappa_2)^2}\,,\label{M-range}
\end{equation}
both arcs can cross the kinematically available range of momenta.
The two resonances are produced with the same $K_z$ but different $K$ and, therefore, they are emitted at different polar angles $\theta_i$ 
given by \eqref{theta-K}.
We obtain a remarkable situation when collision of monochromatic twisted particles at fixed energy
not only produces but also immediately separates several resonances.
In short, twisted particle collision plays the role of a ``built-in mass spectrometer''.

Moreover, these two resonances will not be produced with equal intensity.
One can adjust the collision kinematics in a way
which enhances production of one resonance and suppresses the other.
This is made possible by the presence of interference fringes:
the two resonances can sit in the bright and in the dark fringes.
In fact, this is the situation shown in Fig.~\ref{fig:2d-M1M2}.

\begin{figure}[!htb]
	\centering
	\includegraphics[width=0.307\textwidth]{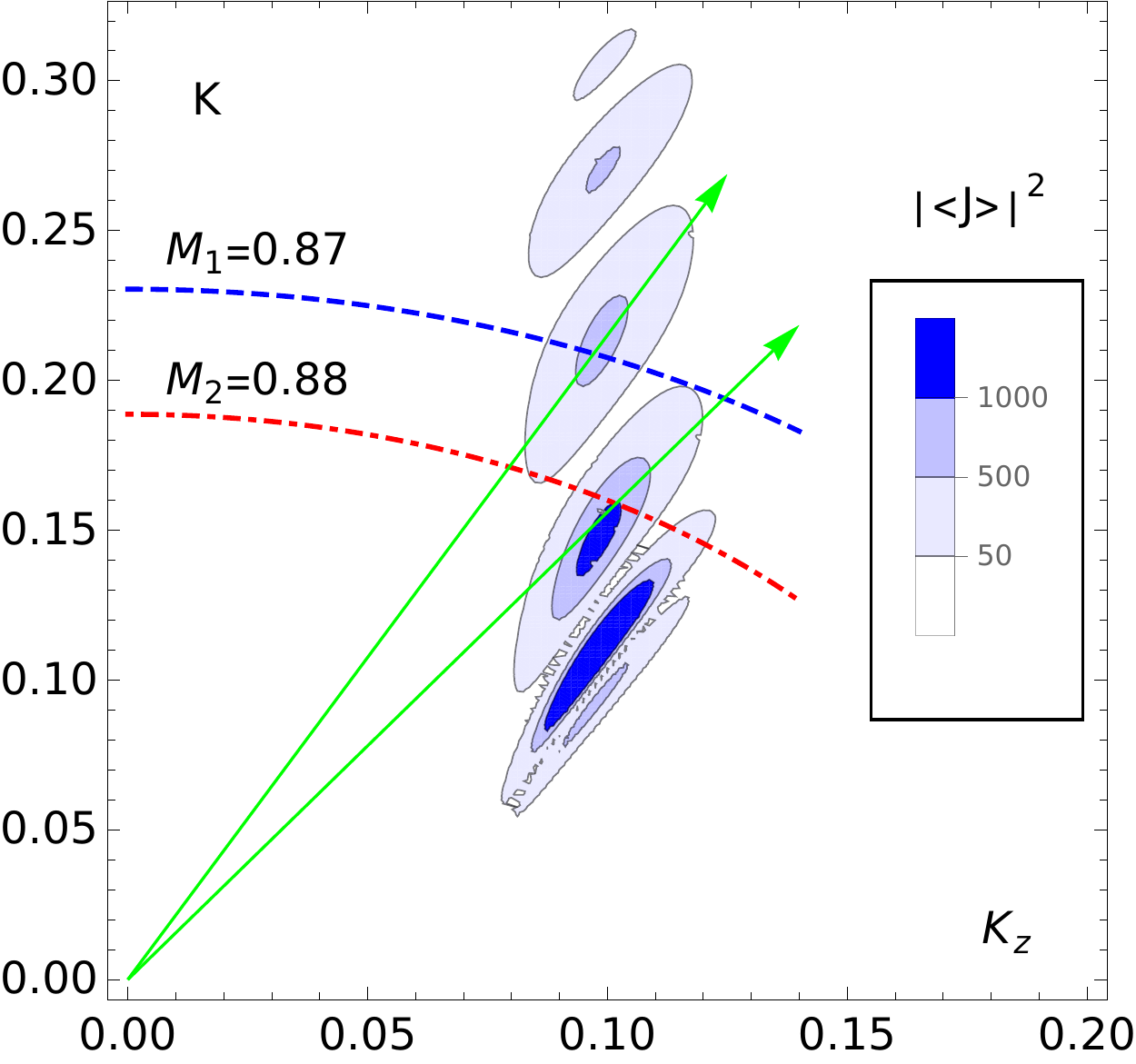}\hspace{5mm}
	\includegraphics[width=0.307\textwidth]{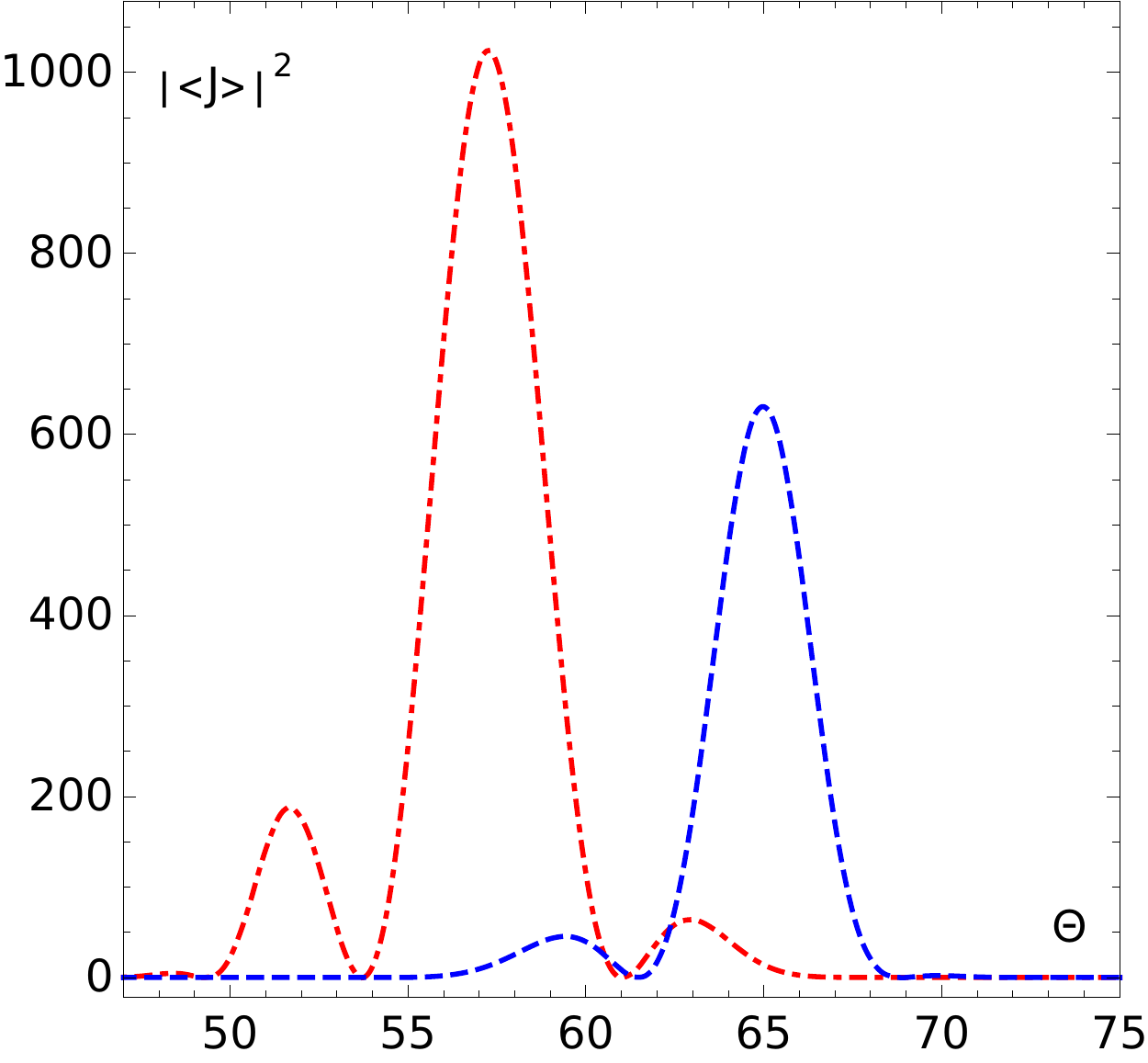}\hspace{5mm}
	\includegraphics[width=0.3\textwidth]{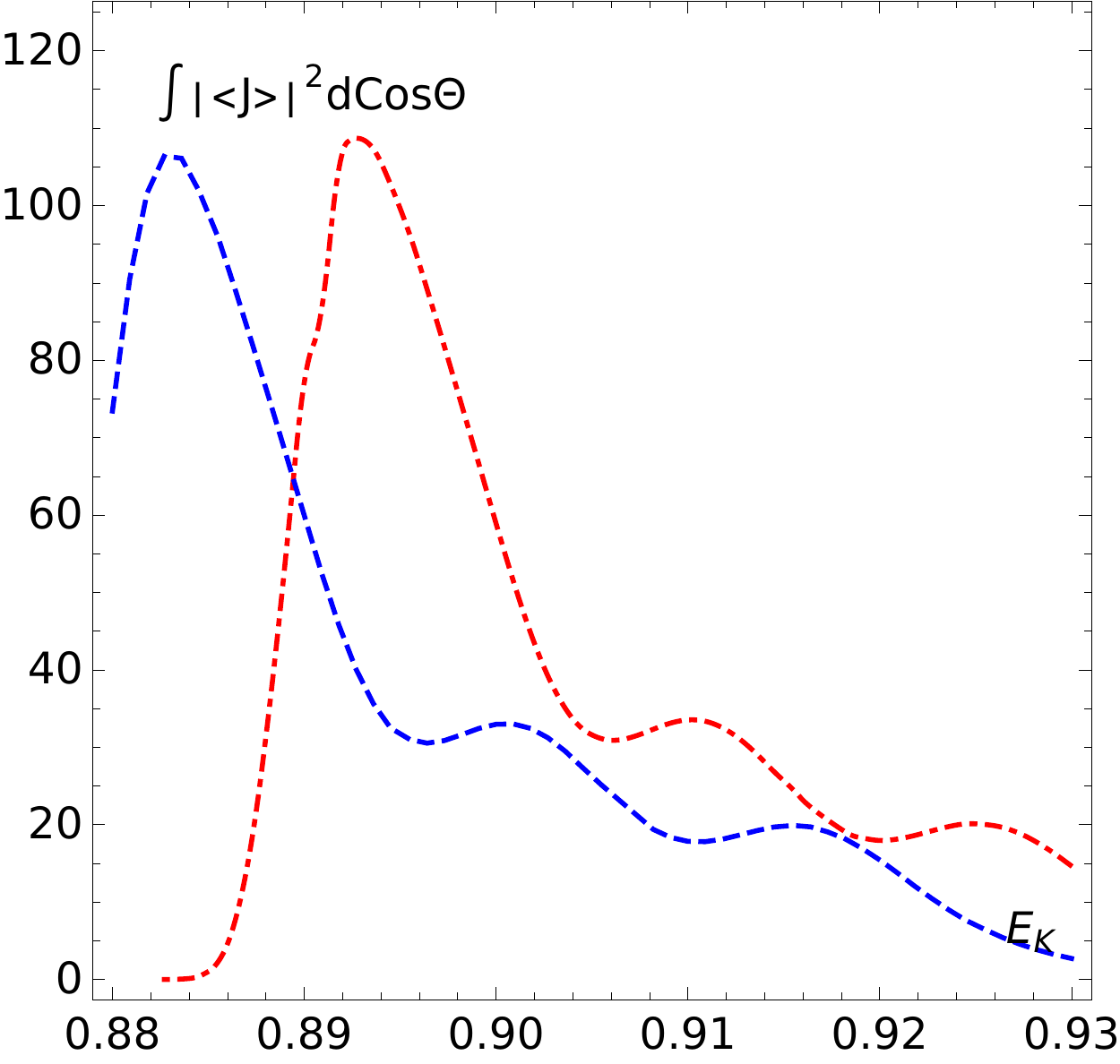}
	\caption{Production of two resonances with masses $M_1 = 0.87$~GeV and $M_2 = 0.88$~GeV 
		in monochromatic twisted particle annihilation with parameters \eqref{smearing-option-2}.
	}
	\label{fig:spectrometry-v0}
\end{figure}

We illustrate this effect in Fig.~\ref{fig:spectrometry-v0}, which demonstrates
production of two narrow resonances with masses $M_1 = 0.87$~GeV and $M_2 = 0.88$~GeV
in collision of monochromatic twisted particles with the following kinematic parameters:
\begin{equation}
\label{smearing-option-2}
E_1=0.48~\GeV, \ E_2=0.42~\GeV, \ \bar{\varkappa}_{1}=0.1~\GeV, \  \bar{\varkappa}_{2}=0.2~\GeV, \ (m_1, m_2) = (5, 1)
\end{equation}
and with $\sigma_i/\bar\varkappa_{i} = 0.1$.
The two arcs in the left plot corresponding to the two resonances with different masses
pass through different fringes.
As a result, one observes remarkably distinct angular distributions of the two resonances,
which is shown in the middle plot.

If one integrates over all angles, one can still clearly separate the two peaks by a scan
over the total collision energy, see right panel of Fig.\ref{fig:spectrometry-v0}. This is, by itself, not a novel feature;
they can, of course, be separated in the usual plane wave annihilation.
What is truly new here with respect to the plane wave case is that 
one can find the energy when the two resonances are  produced simultaneously with 
equal intensity or with any other desired ratio.

A potential benefit of such a simultaneous production setting is that one can reduce or eliminate
a source of systematic uncertainty, which previously seemed unavoidable.
Indeed, if one wants to explore two resonances of distinct masses in the same annihilation process,
one would typically run experiment at one energy and then repeat it at a different energy.
These two runs correspond to different running times and different energies and may involve different systematics.
The proposed scheme avoids that by allowing one to produce two resonances in a single continuously running experiment.

\section{Discussion and conclusions}

Despite several decades of collider experiments, there remains a barely explored opportunity
of colliding particles in initial states which differ from plane waves in an essential way.
One particularly intriguing possibility is to collide twisted particles,
that is, wave packets with helicoidal wave fronts, which carry an adjustable amount of the orbital angular momentum with
respect to the propagation direction. 
With the recent experimental demonstration of twisted electrons \cite{Uchida:2010,Verbeeck:2010,McMorran:2011}
and with suggestions of how to generate GeV range twisted particles \cite{Jentschura:2010ap,Jentschura:2011ih},
such ``twisted collider'' setting seems feasible and promising.

In the recent paper \cite{recent} we argued that the OAM, the new degree of freedom which can be imposed on the initial particles,
leads to several novel kinematic features in the $2\to 1$ annihilation process, which are impossible
to achieve in the usual plane wave collisions. In the present paper, we explored these peculiar features in detail.
The most remarkable one is the intrinsic mass spectrometry, a built-in feature of twisted particle annihilation.
Namely, it is possible to run a monochromatic twisted particle annihilation experiment
at a fixed energy and produce several resonances with close but different masses simultaneously.
These resonances will be produced at different polar angles and can be cleanly separated
by the detector. Moreover, by adjusting the initial particle kinematic, 
one can selectively suppress or enhance production of each of these resonances.
This level of control is unthinkable for the usual plane-wave collisions,
where only the resonance with the mass equal to the total center of mass energy 
of the annihilating particles is produced.

In order to experimentally demonstrate these remarkable features, one needs to overcome
serious technical challenges. First, one needs to bring twisted electrons, photons, or other particles
to the GeV energy range. So far, only electrons with the modest kinetic energy $E_k = 300$ keV have been produced
in electron microscopes \cite{Uchida:2010,Verbeeck:2010,McMorran:2011}.
Second, one needs to achieve sufficiently large transverse momenta of the two colliding particles $\varkappa_i$,
which define the size of the transverse momentum ring \eqref{ring} and the energy range \eqref{E-range}
in which all the remarkable phenomena are displayed.
Third, one must have sufficient control over the transverse localization of the colliding twisted particles.
A twisted state, by definition, must be constructed with respect to a selected axis, the phase nodal line.
The calculations presented in this paper correspond to the ideal alignment when the axes of the two colliding particles coincide.
A large amount of misalignment, either a shift or a tilt, between the two axes would destroy the all-important interference feature.
Slight misalignment can be tolerated, provided the shift is below $1/\varkappa$ and the tilt is much smaller than $\varkappa/k_z$,
see \cite{Ivanov:2016oue}, appendix C. However the exact tolerance limit is not known and 
requires a full 3D treatment of wave packet collision which has not yet been performed.

Overcoming these challenges requires dedicated efforts for instrumentation development.
We believe that the new remarkable opportunities in high energy and hadronic physics offered by twisted particles
represent a sufficiently compelling scientific case and justify further investigation 
into ways to realize such unusual collisions in experiment.

\section*{Acknowledgments}
I.P.I. thanks the Institute of Modern Physics, Lanzhou, China, for financial support and hospitality during his stay.
I.P.I. acknowledges funding from the Portuguese
\textit{Fun\-da\-\c{c}\~{a}o para a Ci\^{e}ncia e a Tecnologia} (FCT) through the FCT Investigator 
contract IF/00989/2014/CP1214/CT0004 and project PTDC/FIS-PAR/29436/2017,
which are partially funded through POCI, COMPETE, Quadro de Refer\^{e}ncia
Estrat\'{e}gica Nacional (QREN), and the European Union.
I.P.I. also acknowledges the support from National Science Center, Poland,
via the project Harmonia (UMO-2015/18/M/ST2/00518).
P.M.Z. is supported by the National Natural Science Foundation of China (Grant No. 11975320)
A.V.P. and N.K. thank the Chinese Academy of Sciences President's International Fellowship Initiative for the support
via grants No. 2019PM0036 (A.V.P.) and No. 2017PM0043 (N.K.).

\appendix

\section{Processes with wave packets}\label{appendix-packet}

The general theory of scattering of non-monochromatic, arbitrarily shaped,
partially coherent beams was developed in \cite{Kotkin:1992bj} in terms of Wigner distribution.
For the specific case of pure, monochromatic, and approximately paraxial initial states this formalism can be simplified 
\cite{Ivanov:2012na,Ivanov:2016oue}.
For the sake of completeness, we review it for the case of $2\to 1$ production.

In the case of plane wave production, when two particles with momenta $\bk_1$ and $\bk_2$
and energies $E_1$ and $E_2$ produce the final particle with energy $E_K$ and momentum $\bK$,
one writes the $S$-matrix amplitude as
\be
S_{PW} = i(2\pi)^4\delta(E_1+E_2-E_K) \delta^{(3)}(\bk_1 + \bk_2 - \bK) {{\cal M}(k_1,k_2;K) \over \sqrt{8 E_1 E_2 E_K}}
\cdot {N_{PW}^3}\,, \label{SPW2}
\ee
where the invariant amplitude ${\cal M}(k_1,k_2;K)$ is calculated according to the standard Feynman rules.
Here, $N_{PW} = 1/\sqrt{V}$ is the plane wave normalization coefficient fixed by the normalization condition
of one particle per large volume $V$. Squaring it, regularizing squares of the delta-functions as
\be
\left[\delta(E_1+E_2-E_K) \delta^{(3)}(\bk_1 + \bk_2 - \bK)\right]^2 = \delta(E_1+E_2-E_K) \delta^{(3)}(\bk_1 + \bk_2 - \bK) 
{V T \over (2\pi)^4}\,,
\ee
integrating over the final phase space $d\Phi_1 = d^3 K/(2\pi)^3$, and dividing by time $T$,
we get the event rate:
\be
\nu = {2 \pi \delta(E_1+E_2-E_K) \over V} {|{\cal M}|^2 \over 8 E_1 E_2 E_K}\,.
\ee
Dividing it by the flux
\be
j = {v \over V}\,, \quad \mbox{where}\quad v = |\bv_1 - \bv_2| = {\sqrt{(k_1k_2)^2-k_1^2k_2^2} \over E_1 E_2}\,,
\label{j-PW}
\ee
where $\bv_1$ and $\bv_2$ are the velocities of the two particles, 
we finally get the cross section:
\be
\sigma = {\nu \over j} = {\pi \delta(E_1+E_2-E_K) \over 4 E_1 E_2 E_K v} |{\cal M}|^2\,.\label{sigma-PW}
\ee

Now we assume that the initial particles are described with the coordinate wave functions
$\psi_1(\br)$ and $\psi_2(\br)$, while the final particle is still a plane wave. 
If the wave function is normalizable, then the normalization condition is $\int d^3 r |\psi(\br)|^2 = 1$, 
where the integral here extends to the entire space.
If it is not, as it is the case for the plane wave and Bessel state,
it goes over a large but finite quantization volume $V$, and the normalization condition
is that one has one particle per volume $V$.
The corresponding momentum-space wave functions are
\be
\varphi(\bk) = \int d^3 r\, \psi(\br)\, e^{-i \bk \br}\,,\quad
\int {d^3 k \over (2\pi)^3} |\varphi(\bk)|^2 = 1\,.
\ee
The $S$-matrix element for the production of final particle with momentum $\bK$ by this initial state 
can be written as
\be
S = \int {d^3 k_1 \over (2\pi	)^3} {d^3 k_2 \over (2\pi)^3} \varphi_1(\bk_1) \varphi_2(\bk_2) S_{PW}\,.
\ee
Since the beams are monochromatic,
the number of scattering events into a given differential volume of the final phase space per unit time is
\be
d\nu = {(2\pi)^7 \delta(E_1+E_2-E_K) \over 4 E_1 E_2} \, |F|^2 \, {d^3 K \over (2\pi)^3 2E_K}\,,\label{dnu1}
\ee
where
\be
F =  \int {d^3 k_1 \over (2\pi)^3} {d^3 k_2 \over (2\pi)^3}  \varphi_1(\bk_1) \varphi_2(\bk_2)\cdot \delta^{(3)}(\bk_1 + \bk_2 - \bK)\, 
\cdot {\cal M}\,.\label{F}
\ee
Note that each $\varphi_i(\bk_i)$ contains a delta-function of the form $\delta(\bk_i^2 + \mu_i^2 - E_i^2)$
because the initial states are monochromatic. 
Thus, the expression for $F$ includes five delta-functions and six integrations and can be represented
as a one-dimensional residual integral. 
Removing the energy delta function in \eqref{dnu1} via integration over $E_K$,
one obtains the following result:
\be
d\nu = {(2\pi)^4 \over 8 E_1 E_2} \, \sqrt{E_K^2-M^2}\, |F|^2 \, { d\Omega_K}\,.\label{dnu2}
\ee
It is known that separation of the event rate into the differential cross section and 
the (conventional) luminosity is uniquely defined only for plane waves and becomes
a matter of convention for non-plane-wave collisions \cite{Kotkin:1992bj}. 
Namely, one needs to adopt a definition of the relative velocity $v$.
One choice is to use the same definition \eqref{j-PW} but with momenta $\bk_i$ replaced
with $\lr{\bk_i}$.
With this definition, the (generalized) cross section for non-plane-wave scattering
can be schematically represented as 
\be
d\sigma = \sigma_0\, R\, d^3 K \,,\quad R \propto |F|^2\,,
\label{ratioR1}
\ee
where $\sigma_0$ is the plane-wave $\bK$-integrated cross section given by \eqref{sigma-PW} and $R$ contains 
factors which are, strictly speaking, process dependent.
There exists a continuous plane-wave limit of these expressions, which was carefully
described in \cite{Kotkin:1992bj,Ivanov:2012na}.
In the plane wave limit, $R \to  \delta^{(3)}(\bk_1 + \bk_2 - \bK)$, and we recover the usual expression.

Transition from the general expressions \eqref{F} and \eqref{dnu2} to the pure Bessel states
can be done with the aid of \eqref{a} and was performed in \cite{Ivanov:2012na}.

\bibliography{res-kinematics-v3.0}
\bibliographystyle{apsrev4-1}

\end{document}